\documentclass[a4paper,11pt]{article}
\usepackage{jcappub} 
\usepackage{multirow}

\title{\huge Halo spin-axis correlations in IllustrisTNG: Lagrangian-to-Eulerian separation mapping, baryonic effects, and redshift-space distortions}

\author[a]{Jun-Sung Moon,}
\author[a,b]{Teppei Okumura}
\author[a,c,d,e,f]{and Ue-Li Pen}
\affiliation[a]{Institute of Astronomy and Astrophysics, Academia Sinica,\\
No. 1, Sec. 4, Roosevelt Rd., Taipei 106319, Taiwan}
\affiliation[b]{Kavli IPMU (WPI), UTIAS, The University of Tokyo,\\
Kashiwa, Chiba 277-8583, Japan}
\affiliation[c]{Canadian Institute for Theoretical Astrophysics, University of Toronto, \\
60 St. George Street, Toronto, Ontario M5S 3H8, Canada}
\affiliation[d]{Dunlap Institute for Astronomy and Astrophysics, University of Toronto, \\
50 St. George Street, Toronto, Ontario M5S 3H4, Canada}
\affiliation[e]{Perimeter Institute for Theoretical Physics, \\
31 Caroline Street North, Waterloo, Ontario N2L 2Y5, Canada}
\affiliation[f]{Canadian Institute for Advanced Research, \\
661 University Avenue, Toronto, Ontario M5G 1M1, Canada}

\emailAdd{jsmoon.astro@gmail.com}

\abstract{
We investigate correlations of halo and stellar spin axes in the TNG300 volume of the IllustrisTNG simulations and compare them with a model based on linear tidal torque theory (LTTT). Dark matter (DM) halo spin axes at $z = 0$ exhibit positive correlations as a function of Lagrangian pair separation. The LTTT-based model describes their separation dependence well, with weak variation in the fitted amplitude across mass thresholds. A Lagrangian-to-Eulerian separation mapping based on the conditional distributions of initial separations at fixed present-day separation, measured directly from the simulations, allows the model to reproduce the overall trends in both real and redshift space without additional fitting. The hydrodynamic and DM-only runs generally show similar DM halo spin-axis correlations, although the DM-only run exhibits a steeper rise at small real-space separations. In the hydrodynamic run, stellar spin-axis correlations are consistent with the same Lagrangian form as their DM halo counterparts, but with approximately 60\% of the DM correlation amplitudes. Positive stellar correlations persist in both real and redshift space, and the mapped model captures their separation dependence. Initial spin-axis correlations of protohalos are substantially stronger than their present-day counterparts at the same Lagrangian separations, consistent with subsequent nonlinear spin evolution weakening these correlations. These results suggest that present-day halo and stellar spin-axis correlations retain signatures of the initial tidal field and that their general behavior can be understood through LTTT-based modeling combined with pair-separation mapping.
}


\begin{document}
\maketitle
\flushbottom

\section{Introduction} \label{sec:intro}

The angular momentum of galaxies is thought to originate from gravitational tidal torques exerted by the surrounding matter distribution during the initial stages of structure formation. Within the framework of linear tidal torque theory (LTTT) \cite{1969ApJ...155..393P, 1970Ap......6..320D, 1984ApJ...286...38W, 1996MNRAS.282..436C}, net angular momentum arises from the misalignment between the shape tensor of a protogalaxy and the local tidal shear field in Lagrangian space. Because the large-scale tidal field itself is spatially correlated, neighboring galaxies can experience correlated tidal torques, which can in turn induce correlations in their angular momentum vectors \cite{2000ApJ...532L...5L, 2000ApJ...543L.107P, 2001ApJ...555..106L}. Spin correlations may provide a useful probe of this coherence imprinted in galaxy angular momenta at early times.

Furthermore, galaxy spins carry directional information that can complement conventional cosmological observables based primarily on the galaxy density field. Spin correlations quantify the spatial coherence of galaxy spin orientations rather than the spatial distribution of galaxies alone. If present-day galaxy spins still retain measurable information about the cosmological initial conditions \cite{2021NatAs...5..283M, 2023ApJ...943..128S, 2023ApJ...952..101M, 2026NatAs..10.1707S}, their correlations offer an additional way to constrain early-Universe physics using low-redshift galaxies, including, for example, tests of primordial parity violation \cite{2020PhRvL.124j1302Y, 2022PhRvD.105h3512M, 2025PhRvL.135n1002S, 2026arXiv260619709Y}.

Theoretical predictions for spin correlations based on LTTT are naturally formulated in Lagrangian space, while spin correlations at late times are measured as a function of the Eulerian separation between galaxies. As gravitational displacements move galaxies away from their initial Lagrangian positions, the initial separation of a galaxy pair does not in general remain the same as its separation at the present epoch \cite{2002MNRAS.332..325P}. It is therefore essential to account for this Lagrangian-to-Eulerian mapping when connecting LTTT-based spin correlations to late-time measurements.

Studies based on cosmological $N$-body simulations have reported spatial correlations between the spins of neighboring dark matter (DM) halos, with the strength and scale varying across different halo samples \cite{2001MNRAS.322..576H, 2002MNRAS.332..325P, 2002A&A...395....1F, 2008ApJ...681..798L, 2013ApJ...762...72T}. In contrast, correlations between galaxy stellar spins have been less explored in hydrodynamic simulations. Ref.~\cite{2015MNRAS.448.3391C} measured correlations between stellar spins at $z = 1.2$ in the Horizon-AGN simulation and found a significant signal for blue and low-mass galaxies. Subsequent work extended the analysis to investigate various forms of spin-related intrinsic alignments \cite{2015MNRAS.454.2736C, 2016MNRAS.461.2702C, 2020MNRAS.492.4268S}, but correlations between spins have received less attention, with grid-locking artifacts in Cartesian grid-based simulations further complicating their interpretation \cite{2015MNRAS.448.3391C, 2015MNRAS.454.2736C}. Cosmological simulations provide access to both the initial conditions and the subsequent evolution, making them well suited for examining the connection between spin correlations measured with respect to Lagrangian and Eulerian separations.

Observational studies have also searched for correlations between galaxy spins. Ref.~\cite{2000ApJ...543L.107P} found evidence for correlated spin axes among nearby galaxies at the $\sim2\sigma$ level. Similar studies using the Sloan Digital Sky Survey detected correlations between spin axes (hereafter spin-axis correlations) at $\sim3\sigma$ significance on scales of a few $h^{-1}\,\mathrm{Mpc}$ \cite{2007ApJ...670L...1L, 2011ApJ...732...99L}. Ref.~\cite{2009MNRAS.392.1225S} reported a hint of $2\,$--$\,3\sigma$ correlation in galaxy spin chirality at separations $\lesssim 0.5\,\mathrm{Mpc}$, based on the sign of the line-of-sight spin component rather than full spin vectors. More recently, ref.~\cite{2025PASJ...77..389T} investigated spin-axis correlations of star-forming galaxies at $z \simeq 1.2$ and $1.5$, finding no significant correlations on scales $\gtrsim 1\,h^{-1}\,\mathrm{Mpc}$, although a weak signal was seen at smaller separations at $z \simeq 1.5$. Despite these efforts, observational measurements necessarily characterize spin correlations using the late-time positions of galaxies, without direct access to their initial Lagrangian separations. Such measurements also remain affected by limitations in sample selection and incomplete observational constraints on the spin directions of galaxies.

A further complication is that observable galaxy spins do not necessarily trace the spins of their DM halos. Stellar spins can be misaligned with those of their host DM halos as angular momentum is redistributed across radii and between different matter components \cite{2010MNRAS.404.1137B, 2015ApJ...812...29T, 2017MNRAS.466.1625Z, 2017MNRAS.472.1163C}. Baryonic processes can also modify halo spin directions through feedback-driven gas flows \cite{2010MNRAS.404.1137B, 2017ApJ...841...16D}, as well as alter the abundance and spatial distribution of halos relative to dark-matter-only (DMO) simulations \cite{2014MNRAS.440.2997V, 2017MNRAS.472.4343C, 2017MNRAS.469.1997D, 2021ApJ...921..112B}. Consequently, it remains unclear how much of the underlying DM halo spin correlation is retained in the stellar component of galaxies.

Even if galaxy spin correlations are understood in real space, observations introduce an additional mapping from real to redshift space. In particular, peculiar velocities distort the inferred pair separations in redshift space, thereby modifying two-point correlations \cite[e.g.,][]{1972MNRAS.156P...1J, 1987MNRAS.227....1K, 1992ApJ...385L...5H, 1998ASSL..231..185H}. Such redshift-space effects have also been studied for intrinsic alignment statistics \cite{2024PhRvD.110f3543M}, where nonlinear redshift-space distortions, such as the Finger-of-God effect, modify galaxy shape correlations \cite{2024PhRvD.109j3501O}. Similarly, spin correlations measured as a function of redshift-space separation may differ from their real-space counterparts \cite{2000ApJ...543L.107P, 2009MNRAS.392.1225S}.

In this work, we use the IllustrisTNG cosmological hydrodynamic and DMO simulations \cite{2018MNRAS.475..624N, 2018MNRAS.475..648P, 2018MNRAS.475..676S, 2018MNRAS.477.1206N, 2018MNRAS.480.5113M} to investigate how signatures of the initial tidal field are reflected in present-day halo and stellar spin-axis correlations. We first measure DM halo spin-axis correlations using present-day spin directions as a function of the initial Lagrangian separations and compare them with an LTTT-based model. We then characterize the mapping between Lagrangian and Eulerian pair separations and test whether it can account for the correlations measured at present-day separations. By comparing the hydrodynamic and DMO runs and extending the analysis to stellar spins, we examine the differences between halo and galaxy spin-axis correlations. We also investigate how redshift-space distortions modify the correlations relevant to observational measurements. Finally, we measure the initial DM spin-axis correlations and compare them with their present-day counterparts at both Lagrangian and Eulerian separations to distinguish the effects of spin reorientation and pair redistribution.

The remainder of this paper is structured as follows. In section~\ref{sec:methods}, we describe the IllustrisTNG simulations, the spin-axis correlation statistics, and the LTTT-based model and separation mapping used in this work. Section~\ref{sec:results} presents our results. We first examine halo spin-axis correlations and their Lagrangian-to-Eulerian mapping in section~\ref{sec:mapping}, followed by redshift-space effects in section~\ref{sec:rsd} and stellar spin-axis correlations in section~\ref{sec:stellar}. We then investigate the initial spin-axis correlations and their relation to the present-day measurements in section~\ref{sec:initial}. Section~\ref{sec:summary} summarizes the main conclusions and discusses the implications of our results. In appendix~\ref{app:veccorr}, we present a supplementary analysis of DM halo spin-vector correlations.

\section{Methods} \label{sec:methods}

\subsection{IllustrisTNG simulations} \label{sec:simulation}

For this study, we use the publicly available snapshot data and subhalo catalogs at $z = 0$, together with the initial conditions at $z = 127$, from TNG300-1 and its DMO counterpart, TNG300-1-Dark, in the IllustrisTNG project \cite{2018MNRAS.475..624N, 2018MNRAS.475..648P, 2018MNRAS.475..676S, 2018MNRAS.477.1206N, 2018MNRAS.480.5113M, 2019ComAC...6....2N}. Both simulations are performed with the moving-mesh code AREPO \cite{2010MNRAS.401..791S} in a periodic cubic box with a comoving side length of $205\,h^{-1}\,\mathrm{Mpc}$. The simulations adopt a flat $\Lambda$CDM cosmology with $\Omega_{\rm m} = 0.3089$, $\Omega_\Lambda = 0.6911$, $\Omega_{\rm b} = 0.0486$, $h = 0.6774$, $\sigma_8 = 0.8159$, and $n_{\rm s} = 0.9667$, consistent with the Planck 2015 results \cite{2016A&A...594A..13P}.

TNG300-1 is a hydrodynamic run with $2500^3$ DM particles and an equal number of initial gas cells. The DM particle mass is $5.9\times10^7\,M_\odot$, and the target baryonic mass resolution is $1.1\times10^7\,M_\odot$. The simulation includes primordial and metal-line radiative cooling, star formation, stellar evolution and chemical enrichment, stellar feedback-driven outflows, supermassive black hole growth, and feedback from accreting black holes \cite{2017MNRAS.465.3291W, 2018MNRAS.473.4077P}. TNG300-1-Dark is evolved from the same initial density field under gravity alone, representing the total matter distribution with $2500^3$ collisionless particles of $7.0\times10^7\,M_\odot$. Comparing these two runs allows us to examine the impact of baryonic processes on spin correlation statistics and their Lagrangian-to-Eulerian separation mapping. 

In both runs, halos are identified using the friends-of-friends (FoF) algorithm, and gravitationally bound substructures within each FoF halo are then identified using the SUBFIND code \cite{2001MNRAS.328..726S, 2009MNRAS.399..497D}. We select subhalos separately in each run according to their total mass, $M_h$, at $z = 0$, including both central and satellite subhalos. We construct three samples with lower mass limits of $10^{11}$, $10^{11.5}$, and $10^{12}\,M_\odot$. In order of increasing mass threshold, the final samples contain 444,393, 152,923, and 52,191 subhalos in the DMO run, and 395,707, 144,571, and 49,173 subhalos in the hydrodynamic run.

For each subhalo, we compute the DM halo spin vector at $z = 0$ from all of its member DM particles,
\begin{equation} \label{eq:spin}
\boldsymbol{L} =
\sum_{\alpha=1}^{n_d} m_d
\left[
(\boldsymbol{x}_{\alpha}-\boldsymbol{x}_c)
\times
(\boldsymbol{v}_{\alpha}-\boldsymbol{v}_c)
\right]\,,
\end{equation}
where $m_d$ is the DM particle mass and $n_d$ is the number of DM particles belonging to the subhalo. Here, $\boldsymbol{x}_{\alpha}$ and $\boldsymbol{v}_{\alpha}$ are the comoving position and peculiar velocity of the $\alpha$th DM particle. The reference position $\boldsymbol{x}_c$ is the location of the particle with the minimum gravitational potential, and $\boldsymbol{v}_c$ is the mass-weighted mean peculiar velocity of all member particles and cells, both taken from the subhalo catalog at $z = 0$. 

Similarly, we compute the stellar spin vector, $\boldsymbol{L}_*$, using only member stellar particles within twice the stellar half-mass radius, $2R_{*,1/2}$, from the subhalo center. We use stellar spins only for subhalos containing more than 300 stellar particles within this aperture. Since this criterion substantially reduces the number of available subhalos in the lowest-mass sample, we exclude that sample from the stellar spin analysis. We therefore consider stellar spins only for the two higher-mass samples, for which the criterion leaves 121,313 and 49,163 subhalos with $M_h > 10^{11.5}\,M_\odot$ and $M_h > 10^{12}\,M_\odot$, respectively.

To determine the initial Lagrangian position of each selected subhalo, we identify its member DM particles at $z = 0$ and trace them back to the initial conditions at $z = 127$ by matching their particle IDs. We define the Lagrangian position, $\boldsymbol{q}$, as the center of mass of these particles in the initial conditions. For each subhalo sample, we use the same $z = 0$ spin directions when measuring correlations as a function of Lagrangian, real-space, or redshift-space separation, changing only the coordinates used to calculate pair separations.

\subsection{Spin-axis correlation statistics} \label{sec:statistics}

To quantify the spatial correlations of subhalo spin orientations, we use the spin-axis correlation function \cite{2000ApJ...543L.107P},
\begin{equation}
\eta =
\left\langle
\left|
\hat{\boldsymbol{L}}^{(1)} \cdot \hat{\boldsymbol{L}}^{(2)}
\right|^2
\right\rangle
-\frac{1}{3}\,,
\end{equation}
where $\hat{\boldsymbol{L}}=\boldsymbol{L}/|\boldsymbol{L}|$ is the unit spin vector, superscripts $(1)$ and $(2)$ identify the two subhalos in a pair, and the average is taken over distinct subhalo pairs within the corresponding separation bin. The subtraction of $1/3$ gives zero expectation for uncorrelated, isotropically distributed three-dimensional spins. Positive and negative values indicate a preference for parallel and perpendicular spin axes, respectively. Because of the squared scalar product, the statistic is invariant under a sign reversal of either spin vector. Unlike the spin-vector correlation function, defined as $\eta_v \equiv \langle \hat{\boldsymbol{L}}^{(1)} \cdot \hat{\boldsymbol{L}}^{(2)} \rangle$, it measures correlations between spin axes without distinguishing parallel from antiparallel spin directions. While we focus on spin-axis correlations in this work, we also examine DM halo spin-vector correlations in appendix~\ref{app:veccorr}.

We measure this statistic as a function of Lagrangian,
real-space Eulerian, or redshift-space Eulerian pair separation, defined in comoving units as
\begin{equation}
r_q = |\boldsymbol{q}^{(1)}-\boldsymbol{q}^{(2)}|\,,
\qquad
r = |\boldsymbol{x}^{(1)}-\boldsymbol{x}^{(2)}|\,,
\qquad
s = |\boldsymbol{s}^{(1)}-\boldsymbol{s}^{(2)}|\,,
\end{equation}
where $\boldsymbol{q}$, $\boldsymbol{x}$, and $\boldsymbol{s}$ denote the initial Lagrangian position, the present-day real-space position, and the present-day redshift-space position of a subhalo, respectively. Subhalo positions in redshift space are obtained by displacing their real-space positions along the line of sight according to their peculiar velocities, while their Lagrangian positions are determined by tracing their bound DM particles back to the initial conditions. Lagrangian and real-space Eulerian separations are computed as the shortest distances between subhalos under periodic boundary conditions. We denote the corresponding correlation functions by $\eta(r_q)$, $\eta(r)$, and $\eta(s)$. Below, the two Eulerian separations are referred to simply as real-space and redshift-space separations.

To estimate the uncertainties, we use spatial jackknife resampling by dividing the simulation volume into $N_{\rm JK} = 5^3$ equal cubic regions, each with a side length of $\sim40\,h^{-1}\,\mathrm{Mpc}$. For each jackknife realization, we exclude one region and recompute the correlation function using only pairs for which both subhalos lie outside the excluded region. The jackknife covariance matrix $\boldsymbol{C}$ is estimated as
\begin{equation}
C_{ij} =
\frac{N_{\rm JK}-1}{N_{\rm JK}}
\sum_{k=1}^{N_{\rm JK}}
\left[\eta_i^{(k)}-\bar{\eta}_i\right]
\left[\eta_j^{(k)}-\bar{\eta}_j\right]\,,
\end{equation}
\begin{equation}
\bar{\eta}_i =
\frac{1}{N_{\rm JK}}
\sum_{k=1}^{N_{\rm JK}}\eta_i^{(k)}\,,
\end{equation}
where $\eta_i^{(k)}$ is the spin-axis correlation function in the $i$th separation bin after excluding the $k$th region, and $\bar{\eta}_i$ is the mean over all jackknife realizations. The error bar in each separation bin is given by $\sigma_{{\rm JK},i} = \sqrt{C_{ii}}$.

\subsection{LTTT-based model and separation mapping} \label{sec:model}

The LTTT-based prediction for spin-axis correlations is formulated in Lagrangian space and is based on the spin--tidal shear correlation \cite{2001ApJ...555..106L}. The second moment of the spin vector $\boldsymbol{L}$ given a unit traceless tidal tensor $\hat{\boldsymbol{T}}$ (normalized such that $\hat{T}_{ij}\hat{T}_{ij}=1$) can be modeled as
\begin{equation} \label{eq:spincov}
\left\langle L_iL_j \mid \hat{\boldsymbol{T}} \right\rangle
=
\frac{1+c}{3}\delta_{ij}
-c\hat{T}_{ik}\hat{T}_{kj}\,,
\end{equation}
where $\langle L_iL_i\mid\hat{\boldsymbol{T}}\rangle=1$. The case of $c = 0$ corresponds to randomly oriented spins, while $c = 1$ recovers the prediction of LTTT for statistically independent inertia and tidal tensors. Using eq.~(\ref{eq:spincov}) as the covariance matrix of a Gaussian conditional distribution of $\boldsymbol{L}$, we integrate over its magnitude to obtain the distribution of unit spin vectors. In the limit $c \ll 1$, we obtain
\begin{equation} \label{eq:spincov_a}
\left\langle \hat{L}_i\hat{L}_j \mid \hat{\boldsymbol{T}} \right\rangle
\simeq
\frac{1+a}{3}\delta_{ij}
-a\hat{T}_{ik}\hat{T}_{kj}\,,
\end{equation}
with $a \equiv 3c/5$ at first order in $c$.\footnote{Retaining higher-order terms in $c$ when deriving eq.~(\ref{eq:spincov_a}) from eq.~(\ref{eq:spincov}) introduces additional tidal tensor terms that cannot generally be absorbed into $a$. Thus, eq.~(\ref{eq:spincov_a}) with $a\equiv3c/5$ is valid only to first order in $c$ for $c\ll1$. We estimate $a$ using a Monte Carlo method by generating $10^6$ random tidal tensors with normally distributed elements, drawing a spin vector for each tensor with covariance given by eq.~(\ref{eq:spincov}), and evaluating the resulting correlation between $\hat{\boldsymbol{L}}$ and $\hat{\boldsymbol{T}}$. The estimated value differs from the first-order approximation by only about 1\% at $c = 0.5$, but reaches $a = 0.69$ at $c = 1$, about 15\% larger than the first-order value of $0.60$, although the exact values of $a$ depend on the assumed distribution of tidal tensors.} The value of $a$ quantifies the strength of the spin--tidal shear correlation \cite{2000ApJ...532L...5L, 2001ApJ...555..106L}, given by $a = 2 - 6\,\langle\sum_i\hat\lambda_i^2\hat L_i^2\rangle$, where $\hat\lambda_i$ are the eigenvalues of $\hat{\boldsymbol{T}}$ and $\hat L_i$ are the components of the unit spin vector along the corresponding eigenvectors.

If the inertia tensors at different positions are statistically independent given the tidal tensors, the spin-axis correlation can be expressed in terms of the initial tidal field:
\begin{equation}
\begin{aligned}
\eta(r_q)
&\simeq
\left\langle
\left(\frac{1+a}{3}\delta_{ij}
-a\hat{T}_{ik}\hat{T}_{kj}\right)
\left(\frac{1+a}{3}\delta_{ij}
-a\hat{T}'_{il}\hat{T}'_{lj}\right)
\right\rangle
-\frac{1}{3}
\\
&=
a^2
\left[
\left\langle
\hat{T}_{ik}\hat{T}_{kj}
\hat{T}'_{il}\hat{T}'_{lj}
\right\rangle
-\frac{1}{3}
\right]\,,
\end{aligned}
\end{equation}
where $\hat{\boldsymbol{T}}$ and $\hat{\boldsymbol{T}}'$ are evaluated at the initial Lagrangian positions $\boldsymbol{q}^{(1)}$ and $\boldsymbol{q}^{(2)}$, respectively. Following the derivation in ref.~\cite{2001ApJ...555..106L}, applying Wick's theorem and assuming a power-law density correlation lead to
\begin{equation} \label{eq:fit3d}
\eta(r_q)
=
\frac{a^2}{6}\,
\frac{\xi_R^2(r_q;R_f)}{\xi_R^2(0;R_f)}\,.
\end{equation}
Here, $\xi_R(r_q;R_f)$ is the correlation function of the initial density field smoothed on a scale $R_f$ to account for the finite extent of protohalos. Given the linear matter power spectrum $P_\delta(k)$ and a spherical top-hat window $W(k,R_f)$,
\begin{equation}
\xi_R(r_q;R_f)
=
\frac{1}{2\pi^2}
\int_0^\infty
k^2P_\delta(k)
W^2(k,R_f)
\frac{\sin(kr_q)}{kr_q}\,\mathrm{d}k\,.
\end{equation}
In this study, we adopt a power-law matter power spectrum, $P_\delta(k) \propto k^{-2}$. We use eq.~(\ref{eq:fit3d}) to model the correlations of the present-day spins measured as a function of $r_q$, treating $a$ as an effective fitting parameter rather than a direct measure of the spin--tidal shear correlation. When fitting the present-day spin-axis correlations, we fix $R_f$ to the mean Lagrangian radius, $\langle R_L \rangle$, for each mass sample in each run, with $R_L =[3M_h/(4\pi\bar{\rho}_m)]^{1/3}$, where $\bar{\rho}_m$ is the mean comoving matter density. In order of increasing mass, the adopted values of $R_f$ are 0.89, 1.28, and 1.84 $h^{-1}\,\mathrm{Mpc}$ in the DMO run, and 0.89, 1.27, and 1.81 $h^{-1}\,\mathrm{Mpc}$ in the hydrodynamic run. For the stellar spin samples, the corresponding values are 1.34 and 1.81 $h^{-1}\,\mathrm{Mpc}$ for $M_h > 10^{11.5}\,M_\odot$ and $M_h > 10^{12}\,M_\odot$, respectively. For the initial spin-axis correlations examined in section~\ref{sec:initial}, we instead allow both $a$ and $R_f$ to vary to better reproduce their separation dependence.

To relate the LTTT-based model to real-space measurements, we construct a separation mapping kernel, $\boldsymbol{K} = (K_{ij})$, for each sample,
\begin{equation} \label{eq:sep_kernel}
K_{ij}
\equiv
P(r_{q,i} \mid r_j)
=
\frac{N_{ij}}{N_j}\,,
\end{equation}
where $N_{ij}$ is the number of pairs that belong to the $i$th Lagrangian bin and the $j$th real-space bin, with $N_j = \sum_i N_{ij}$ and $\sum_i K_{ij} = 1$. Thus, $P(r_{q,i} \mid r_j)$ denotes the probability that a pair in the $j$th real-space bin belongs to the $i$th Lagrangian bin. Assuming no additional dependence of the spin-axis correlation on real-space separation at fixed Lagrangian separation, the mapped model at each real-space separation bin $r_j$ is given by
\begin{equation} \label{eq:mapped_model}
\eta(r_j) =
\sum_i K_{ij}\,
\eta(r_{q,i};a,R_f)\,.
\end{equation}

We first fit eq.~(\ref{eq:fit3d}) to the measured $\eta(r_q)$ to obtain the best-fit value $a_L$. We then set $a = a_L$ in eq.~(\ref{eq:mapped_model}) to obtain a real-space prediction, without any additional fitting to the measured $\eta(r)$. We also fit the mapped model directly to the measured $\eta(r)$, allowing $a$ to vary to obtain a separate best-fit value, $a_E$. The smoothing scale $R_f$ and the separation kernel are held fixed in this fit. Both fits use $\chi^2$ minimization with the corresponding full jackknife covariance matrix $\boldsymbol{C}$. We use 11 logarithmic separation bins of width $0.25$ dex, spanning $-1.25 \leq \log(r/h^{-1}\,\mathrm{Mpc}) \leq 1.50$. Only bins containing more than 100 pairs are included in the fits. We do not impose the bounds on $a$ in either fit. The uncertainties in the fitted parameters are estimated as the square roots of the diagonal elements of the parameter covariance matrix. The mapped model for the redshift-space correlation, $\eta(s)$, is constructed analogously using the corresponding mapping kernel $\boldsymbol{K}_s = (K_{s,ij})$, where $K_{s,ij} = P(r_{q,i} \mid s_j)$.

\section{Results} \label{sec:results}

In this section, we present the spin-axis correlations measured in the TNG300 simulations. In section~\ref{sec:mapping}, we first examine the present-day DM halo spin-axis correlations as functions of Lagrangian and real-space Eulerian separations, testing the LTTT-based model and the separation mapping. We then extend the DM halo spin analysis to redshift space in section~\ref{sec:rsd}. In section~\ref{sec:stellar}, we examine stellar spin-axis correlations as functions of Lagrangian, real-space, and redshift-space separations. In section~\ref{sec:initial}, we compare the initial DM halo spin-axis correlations with their present-day counterparts to examine the effects of late-time evolution.

\subsection{Halo spin-axis correlations and Lagrangian-to-Eulerian mapping} \label{sec:mapping}

\begin{figure}[tp]
\centering
\includegraphics[width=.98\textwidth]{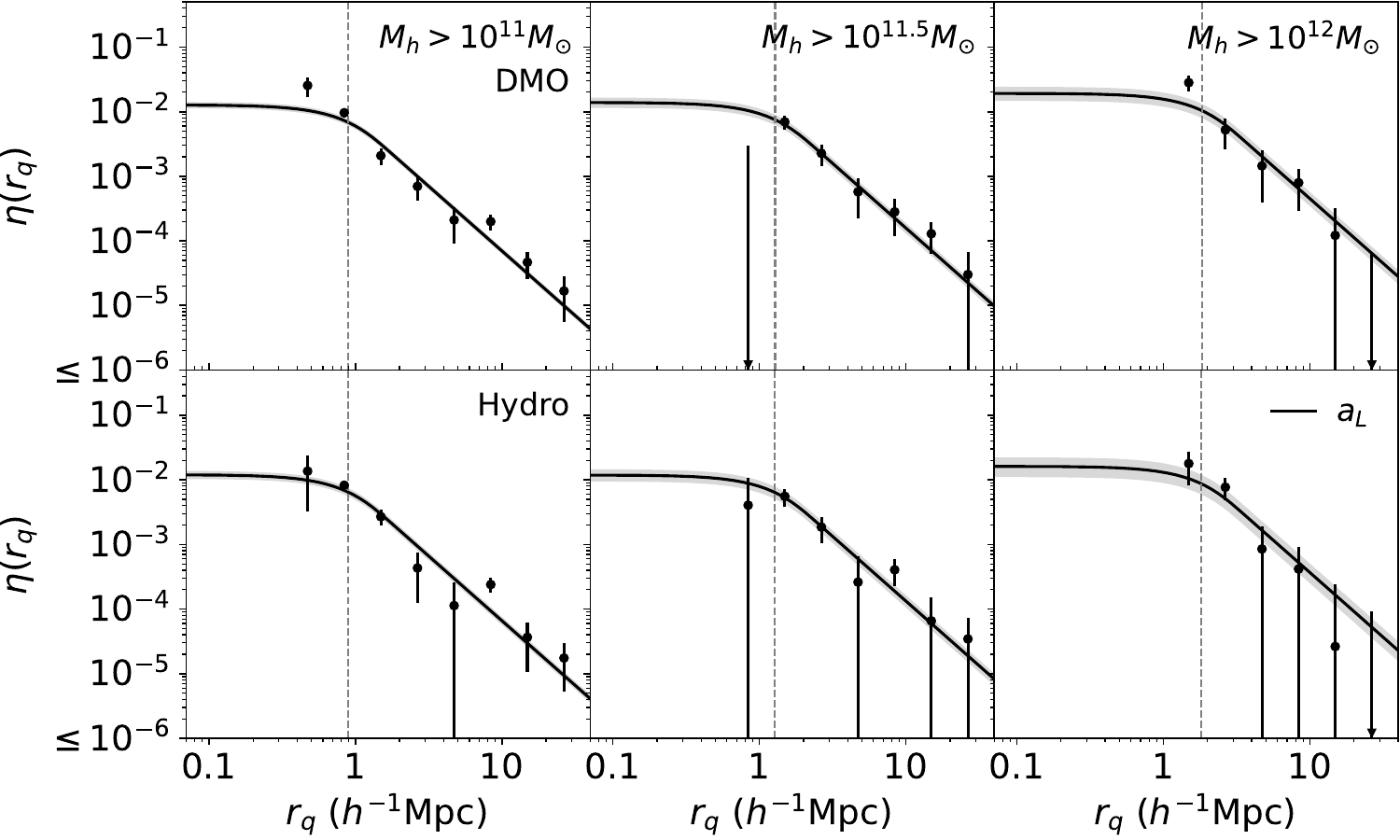}
\caption{DM halo spin-axis correlation functions, $\eta(r_q)$, as a function of Lagrangian separation $r_q$. Spin directions are measured at $z = 0$, while pair separations are calculated from the initial positions of the subhalos. Columns show samples with lower mass limits of $10^{11}$, $10^{11.5}$, and $10^{12}\,M_\odot$, from left to right. The top and bottom rows correspond to the DMO and hydrodynamic runs, respectively. Solid curves show the best-fit LTTT-based model given by eq.~(\ref{eq:fit3d}), with amplitude $a_L$ and smoothing scale $R_f$ fixed to the mean Lagrangian radius of each sample (vertical dashed lines). In this and subsequent correlation plots, only bins containing more than 100 pairs are shown, and error bars indicate jackknife uncertainties. Downward arrows mark measurements below $10^{-6}$. Shaded regions indicate the $1\sigma$ uncertainties of the fitted models. \label{fig:f1}}
\end{figure}

Figure~\ref{fig:f1} shows the DM halo spin-axis correlation functions, $\eta(r_q)$, measured using the present-day spin directions and the initial Lagrangian separations of the subhalo pairs. In the DMO run, shown in the top row, the correlations exhibit an overall positive signal that decreases with increasing $r_q$ in all three samples with different mass thresholds. The correlation amplitudes show a slight tendency to increase toward higher mass thresholds, although they remain statistically consistent within their uncertainties. The LTTT-based model given by eq.~(\ref{eq:fit3d}), with $R_f$ fixed to the mean Lagrangian radius, captures the overall behavior of the measured spin-axis correlations across the three samples. The best-fit parameters and corresponding reduced chi-square values are summarized in table~\ref{tab:t1}. For the samples with lower mass limits of $10^{11}$, $10^{11.5}$, and $10^{12}\,M_\odot$, the best-fit values are $a_L = 0.28\,\pm\,0.02$, $0.29\,\pm\,0.03$, and $0.34\,\pm\,0.04$, respectively, with corresponding reduced chi-square values of $2.22$, $0.89$, and $1.27$. Our $a_L$ provides an effective estimate of the spin--tidal shear correlation strength through fits to the spin-axis correlations. The inferred values are comparable in magnitude to $a = 0.24\,\pm\,0.02$ measured directly from spin--tidal shear correlations in an $N$-body simulation by ref.~\cite{2000ApJ...532L...5L}, but substantially larger than $a = 0.07\,\pm\,0.01$ obtained from a higher-resolution simulation by ref.~\cite{2002MNRAS.332..339P}.

\begin{table}[tp]
\centering
\begin{tabular}{ccccccc}
\hline
& $M_{h,\min}$ & $R_f$ & $a_L$ & $\chi_L^2/\mathrm{dof}$ & $a_E$ & $\chi_E^2/\mathrm{dof}$ \\
\hline

& $10^{11} $ & 0.89 & 0.28 $\pm$ 0.02 & 2.22 & 0.32 $\pm$ 0.03 & 4.18 \\
DMO & $10^{11.5} $ & 1.28 & 0.29 $\pm$ 0.03 & 0.89 & 0.38 $\pm$ 0.04 & 2.48 \\
& $10^{12} $ & 1.84 & 0.34 $\pm$ 0.04 & 1.27 & 0.32 $\pm$ 0.06 & 1.64 \\
\hline
& $10^{11} $ & 0.89 & 0.27 $\pm$ 0.02 & 2.21 & 0.30 $\pm$ 0.03 & 0.90 \\
Hydro & $10^{11.5} $ & 1.27 & 0.27 $\pm$ 0.03 & 0.53 & 0.32 $\pm$ 0.04 & 0.82 \\
& $10^{12} $ & 1.81 & 0.31 $\pm$ 0.05 & 0.64 & 0.24 $\pm$ 0.09 & 1.30 \\
\hline

\end{tabular}
\caption{Best-fit parameters for the DM halo spin-axis correlations shown in figures~\ref{fig:f1} and \ref{fig:f3}. The parameters $a_L$ and $a_E$ are obtained by fitting $\eta(r_q)$ and $\eta(r)$ using eqs.~(\ref{eq:fit3d}) and (\ref{eq:mapped_model}), respectively. The corresponding reduced chi-square values are $\chi_L^2/\mathrm{dof}$ and $\chi_E^2/\mathrm{dof}$, where dof denotes the number of degrees of freedom. In this and subsequent tables, $M_{h,\min}$ represents the lower subhalo mass limit, and all uncertainties are quoted at the $1\sigma$ level. $M_{h,\min}$ and $R_f$ are given in units of $M_\odot$ and $h^{-1}\,\mathrm{Mpc}$, respectively.} \label{tab:t1}
\end{table}

The spin-axis correlations in the hydrodynamic run, shown in the bottom row of figure~\ref{fig:f1}, exhibit similar trends with Lagrangian separation and mass threshold. The best-fit values are $a_L = 0.27\,\pm\,0.02$, $0.27\,\pm\,0.03$, and $0.31\,\pm\,0.05$ for the three samples, respectively. The corresponding reduced chi-square values are $2.21$, $0.53$, and $0.64$. The similarity between the DMO and hydrodynamic runs indicates that the overall dependence of the present-day DM spin-axis correlation on Lagrangian separation is robust to the inclusion of baryonic physics. We stress that, although the pair separations in figure~\ref{fig:f1} are defined from the initial conditions, the spin directions are measured at $z = 0$. Since nonlinear evolution can alter individual spin directions, the measured $\eta(r_q)$ should not be interpreted as the spin-axis correlation at the initial epoch \cite{2002MNRAS.332..325P}. Rather, our comparison with the LTTT-based model tests whether present-day spin-axis correlations reflect the dependence on Lagrangian separation expected from the spatial coherence of the initial tidal field \cite{2000ApJ...532L...5L, 2001ApJ...555..106L}. The broad agreement suggests that the LTTT-based model provides a useful approximate description of the present-day spin-axis correlations when expressed as a function of initial Lagrangian separations, despite the nonlinear evolution of individual spin directions. We examine the initial spin-axis correlations in section~\ref{sec:initial} to investigate how they differ from their present-day counterparts.

\begin{figure}[tp]
\centering
\includegraphics[width=\textwidth]{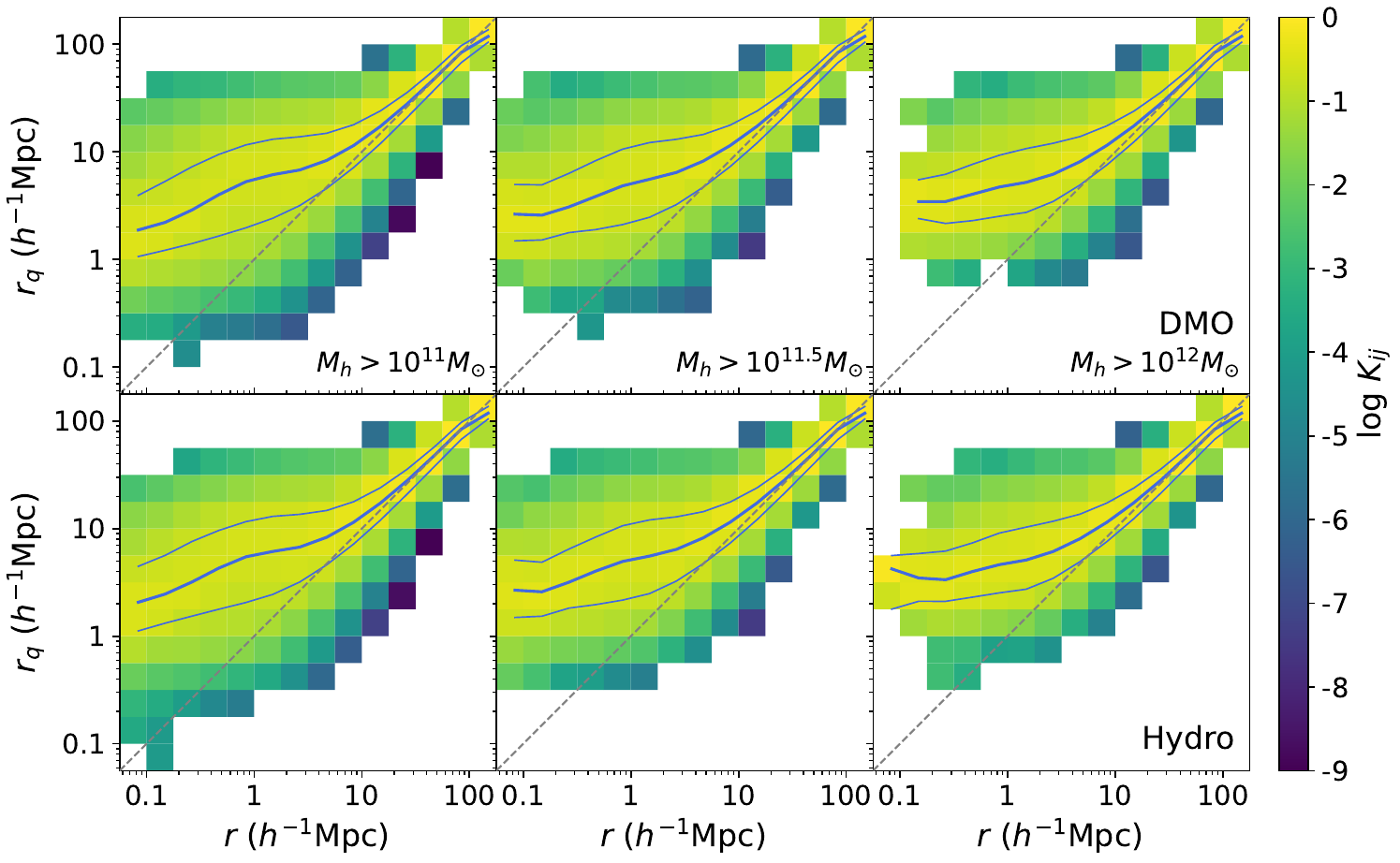}
\caption{Separation mapping kernel $\boldsymbol{K}$ from Lagrangian to real-space Eulerian separations. The sample in each panel is identical to that in the corresponding panel of figure~\ref{fig:f1}. Colors show $\log K_{ij}$, where $K_{ij} = P(r_{q,i} \mid r_j)$ is the conditional probability of Lagrangian separation at fixed real-space separation, defined in eq.~(\ref{eq:sep_kernel}). Thick and thin blue curves indicate the median and the 16th and 84th percentiles of $r_q$ at fixed $r$, respectively. White bins contain no pairs, and diagonal dashed lines mark $r_q = r$. \label{fig:f2}}
\end{figure}

We next test whether mapping these best-fit Lagrangian models to real-space Eulerian separations can reproduce the measured $\eta(r)$ at $z = 0$, following the procedure described in section~\ref{sec:model}. Figure~\ref{fig:f2} shows the Lagrangian-to-real-space separation mapping kernel, $\boldsymbol{K}$, for the same subhalo samples as in figure~\ref{fig:f1}. At large real-space separations, the median Lagrangian separation approaches $r$, and the conditional distribution is concentrated near the diagonal line $r_q = r$. At small $r$, however, a substantial fraction of pairs had initial separations much larger than their present-day separations. For example, pairs at $r = 1\,h^{-1}\,\mathrm{Mpc}$ have a median Lagrangian separation of approximately $5\,h^{-1}\,\mathrm{Mpc}$ in all six samples. The broad distribution of $r_q$ at fixed $r$ also shows that a single real-space separation bin contains pairs spanning a wide range of initial separations. Differences among the samples are noticeable only at small $r$, where the distributions for the higher-mass samples are less extended toward small $r_q$. The DMO and hydrodynamic runs exhibit similar Lagrangian-to-real-space separation mappings for samples with the same mass thresholds.

\begin{figure}[tp]
\centering
\includegraphics[width=.98\textwidth]{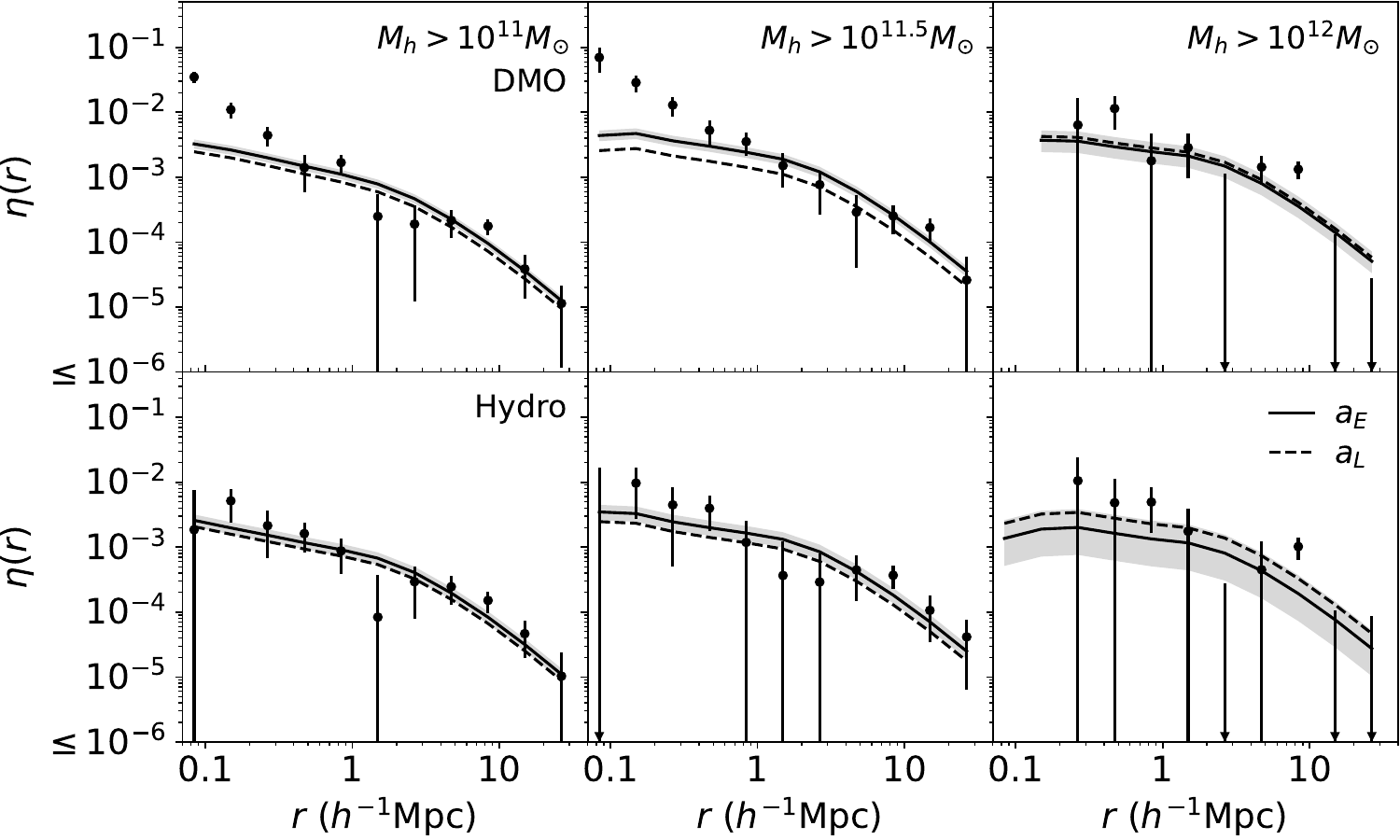}
\caption{DM halo spin-axis correlation functions, $\eta(r)$, as a function of real-space Eulerian separation $r$ at $z = 0$. The sample in each panel is identical to that in the corresponding panel of figure~\ref{fig:f1}. Solid curves show the mapped model given by eq.~(\ref{eq:mapped_model}), fitted directly to $\eta(r)$ with amplitude $a_E$. Dashed curves show predictions obtained by mapping the best-fit Lagrangian models with amplitude $a_L$, without refitting to the real-space measurements. \label{fig:f3}}
\end{figure}

Figure~\ref{fig:f3} shows the DM halo spin-axis correlation functions, $\eta(r)$, measured as a function of real-space Eulerian separation at $z = 0$. Dashed curves show the predictions obtained by mapping the best-fit Lagrangian models from figure~\ref{fig:f1}, using $a = a_L$ without additional fitting. Solid curves show the mapped model fitted directly to the real-space measurements, with amplitude $a_E$. The mapped predictions with $a_L$ reproduce the overall behavior of the measured real-space correlations. The shapes of the mapped models can be understood from the Lagrangian correlations in figure~\ref{fig:f1} and the separation mapping in figure~\ref{fig:f2}. At small $r$, the conditional distribution of $r_q$ extends over a broad range and varies only weakly with $r$. At the same time, the LTTT-based model predicts nearly constant correlations for $r_q \lesssim 1\,h^{-1}\,\mathrm{Mpc}$. Averaging the Lagrangian model over the conditional distribution therefore produces mapped correlations that vary relatively weakly with $r$ at small separations. At larger $r$, the distribution becomes concentrated near $r_q=r$, and the mapped correlations more closely follow the Lagrangian model, declining more rapidly with separation.

A closer look at figure~\ref{fig:f2} reveals a flattening of the median $r_q$ around $r = 1\,$--$\,3\,h^{-1}\,\mathrm{Mpc}$ before it decreases more rapidly toward smaller $r$. This feature may reflect a transition at these scales in the relative contributions of one-halo and two-halo pairs. However, this flattening of the median does not necessarily produce a corresponding feature in the mapped correlations, because $\eta(r_q)$ decreases rapidly with increasing $r_q$, making the mapped correlations more sensitive to the low-$r_q$ tail of the conditional distribution than to its median alone. Interestingly, the measured $\eta(r)$ shows a hint of a dip around the same scales in some panels of figure~\ref{fig:f3}, particularly for the lower mass thresholds, which is not reproduced by the mapped models.

In the DMO run, the mapped predictions reproduce the overall decrease in the correlation amplitude with increasing $r$, but show substantial discrepancies at $r \leq 0.3\,h^{-1}\,\mathrm{Mpc}$, particularly for the two lower mass-threshold samples. The measured correlations are substantially stronger than the predictions in the innermost bins, and this discrepancy remains even after fitting $a_E$. The best-fit values are $a_E = 0.32\,\pm\,0.03$, $0.38\,\pm\,0.04$, and $0.32\,\pm\,0.06$ for the three samples, respectively, with corresponding reduced chi-square values of $4.18$, $2.48$, and $1.64$, as summarized in table~\ref{tab:t1}. The excess correlations at small separations may arise from one-halo contributions, where nonlinear tidal interactions among nearby halos can induce additional spin alignments. Previous simulation studies have shown that such interactions can strengthen the alignments between halo spins and the orbital angular momentum of interacting halo pairs \cite{2017MNRAS.466.4875L, 2021ApJ...909...34M, 2021ApJ...914...86A, 2026ApJ..1004..180O}. Such interaction-driven alignments, which are not explicitly accounted for by the LTTT-based model, introduce an additional dependence of the spin-axis correlation on real-space separation at fixed Lagrangian separation.

In the hydrodynamic run, the mapped predictions show good agreement with the measured real-space correlations. The best-fit values are $a_E = 0.30\,\pm\,0.03$, $0.32\,\pm\,0.04$, and $0.24\,\pm\,0.09$, with corresponding reduced chi-square values of $0.90$, $0.82$, and $1.30$. The fitted real-space amplitudes $a_E$ are comparable to their Lagrangian counterparts $a_L$ at the $\sim\!1\sigma$ level in all three samples. The agreement is generally better than in the DMO run, particularly at small separations. These results suggest that the Lagrangian-to-real-space separation mapping captures the general separation dependence of the present-day DM spin-axis correlations.

Overall, the comparison between the DMO and hydrodynamic runs suggests that baryonic physics has little impact on the Lagrangian spin-axis correlations, their best-fit amplitudes $a_L$, or the Lagrangian-to-real-space separation mapping. Their differences are most apparent in the real-space spin-axis correlations at small separations, where the excess seen in the DMO run is less pronounced in the hydrodynamic run. This difference may be related to baryonic processes that modify halo spin directions through feedback-driven gas flows and alter the spatial distribution of subhalos \cite[e.g.,][]{2010MNRAS.404.1137B, 2014MNRAS.440.2997V, 2017MNRAS.472.4343C}. Such effects may suppress the development of strong interaction-driven spin alignments at small separations, reducing the excess correlations seen in the DMO run.

Previous observational studies compared galaxy spin-axis correlations with LTTT-based predictions without explicitly accounting for the Lagrangian-to-Eulerian mapping of pair separations \cite{2000ApJ...543L.107P, 2007ApJ...670L...1L, 2011ApJ...732...99L, 2025PASJ...77..389T}. Here, we instead use the full conditional distribution of Lagrangian separations at fixed Eulerian separation, measured directly from the simulations, to construct the mapped predictions. The overall agreement between the mapped predictions and the Eulerian measurements suggests that the nonlinear redistribution of pair separations can account for much of the difference between spin-axis correlations expressed as functions of Lagrangian and Eulerian separations.

\subsection{Redshift-space distortion effects} \label{sec:rsd}

We now extend the analysis of section~\ref{sec:mapping} to redshift space, where peculiar velocities cause redshift-space distortions by modifying the inferred pair separations along the line of sight. We retain the same $z = 0$ spin directions and measure their correlations as a function of redshift-space Eulerian separation, $s$. Using the corresponding mapping kernel $\boldsymbol{K}_s$, with $K_{s,ij} = P(r_{q,i} \mid s_j)$, we first obtain predictions for $\eta(s)$ from the best-fit Lagrangian models using $a_L$ without refitting. We then fit the mapped model directly to the redshift-space measurements, allowing the amplitude to vary while holding $R_f$ and $\boldsymbol{K}_s$ fixed. We denote the resulting best-fit amplitude by $a_s$.

We compute redshift-space separations $s$ by taking each Cartesian axis of the simulation box in turn as the line of sight. For a pair of subhalos, the redshift-space separation is
\begin{equation}
s = \left|
\boldsymbol{r}
+\frac{\Delta\boldsymbol{v}\cdot\hat{\boldsymbol{n}}}{H_0}
\hat{\boldsymbol{n}}
\right|\,,
\end{equation}
where $\boldsymbol{r}$ is the real-space pair separation vector, $\Delta\boldsymbol{v}$ is the relative peculiar velocity, $\hat{\boldsymbol{n}}$ is the unit vector along the chosen line of sight, and $H_0$ is the present-day Hubble parameter. We measure the spin-axis correlation separately for the three line-of-sight directions and average them, $\eta(s) = [\eta_x(s)+\eta_y(s)+\eta_z(s)]/3$. The jackknife covariance is calculated from the values of $\eta(s)$ obtained in the individual jackknife realizations, using regions defined from the real-space positions. We also construct the mapping kernel $\boldsymbol{K}_s$ by counting pairs from all three line-of-sight directions in each $(r_q,s)$ bin and normalizing the counts at fixed $s$.

\begin{figure}[tp]
\centering
\includegraphics[width=\textwidth]{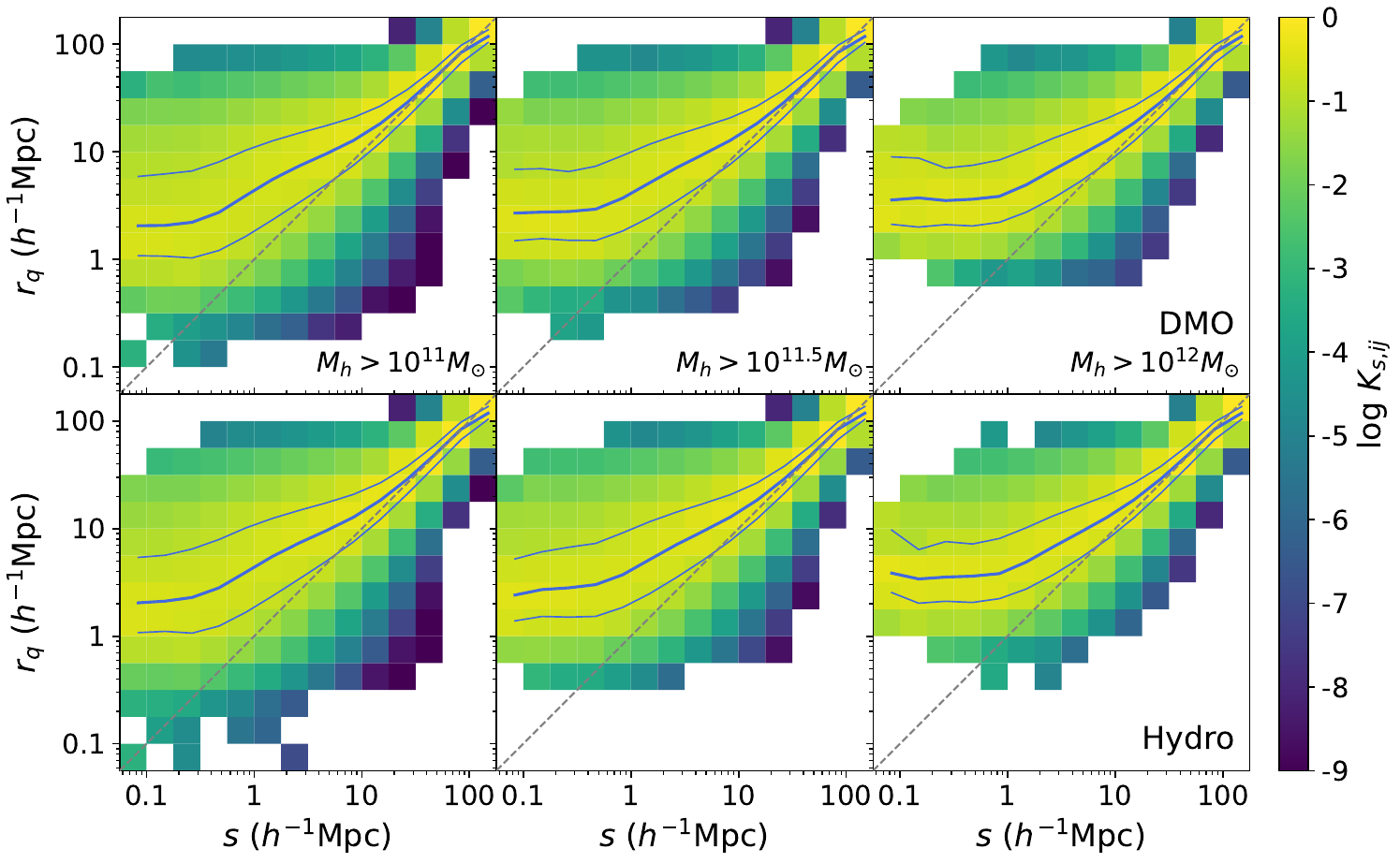}
\caption{Same as figure~\ref{fig:f2}, but for the separation mapping kernel $\boldsymbol{K}_s$ from Lagrangian to redshift-space Eulerian separations. Diagonal dashed lines mark $r_q = s$. \label{fig:f4}}
\end{figure}

Figure~\ref{fig:f4} shows the Lagrangian-to-redshift-space separation mapping kernels for the same subhalo samples as in figure~\ref{fig:f2}. The mapping kernels show qualitatively similar features to their real-space counterparts. Deviations from the diagonal relation $r_q = s$ are most pronounced at small $s$, where each redshift-space separation bin contains pairs drawn from a broad range of initial Lagrangian separations. In the smallest separation bins, the median $r_q$ is approximately $2\,$--$\,4\,h^{-1}\,\mathrm{Mpc}$ and increases with the mass threshold. At small separations, $s < 1\,h^{-1}\,\mathrm{Mpc}$, the median $r_q$ varies relatively weakly with $s$, producing a more pronounced plateau than in the real-space mapping shown in figure~\ref{fig:f2}. This plateau is most evident for the highest-mass sample. Toward larger $s$, the mapping approaches the diagonal relation, and the redshift-space separation more closely traces the initial separation. Compared with the real-space mapping, the redshift-space distributions exhibit more extended tails, particularly toward small $r_q$ at large $s$. This behavior may reflect redshift-space distortions such as the Finger-of-God effect, which arises from velocity dispersion in virialized regions. These trends are similar in the DMO and hydrodynamic runs across all three mass thresholds.

\begin{figure}[tp]
\centering
\includegraphics[width=.98\textwidth]{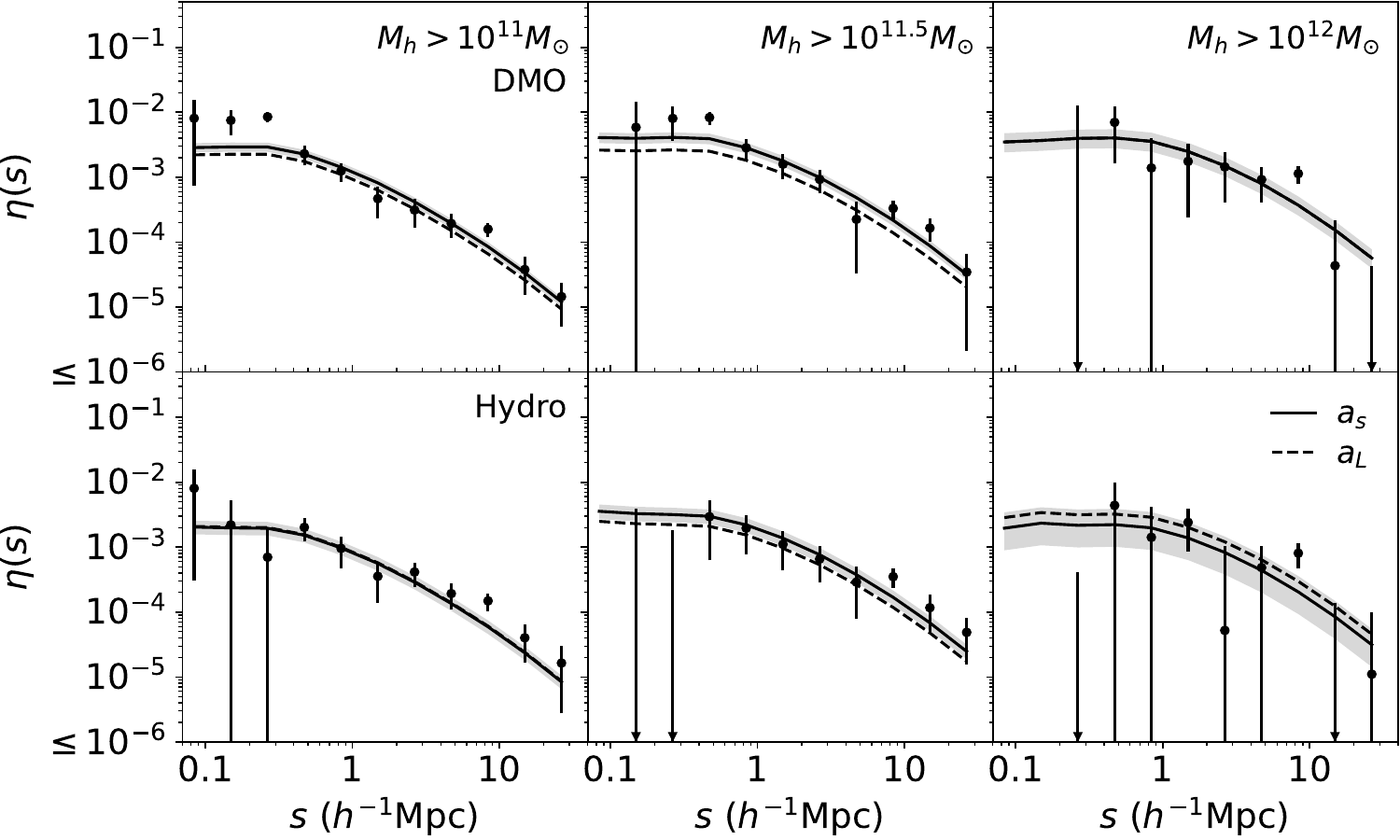}
\caption{Same as figure~\ref{fig:f3}, but for $\eta(s)$ as a function of redshift-space Eulerian separation $s$. The amplitude obtained by fitting directly to $\eta(s)$ is denoted by $a_s$. \label{fig:f5}}
\end{figure}

Figure~\ref{fig:f5} compares the measured DM halo spin-axis correlations in redshift space with the mapped models. Dashed curves show predictions based on the Lagrangian amplitudes $a_L$, while solid curves show direct fits with amplitudes $a_s$. The fitting results for the redshift-space measurements are summarized in table~\ref{tab:t2}. For the DMO samples in the top row, the models generated using $a_L$ and the mapping kernel $\boldsymbol{K}_s$ capture the overall decline of $\eta(s)$ with increasing separation well. In order of increasing mass threshold, the best-fit amplitudes are $a_s = 0.32\,\pm\,0.03$, $0.36\,\pm\,0.03$, and $0.34\,\pm\,0.06$, with reduced chi-square values of $2.54$, $1.15$, and $1.08$, respectively. Refitting raises the model amplitudes for the two lower mass thresholds, while the prediction and the refitted model nearly coincide for the highest threshold. Although less pronounced than in figure~\ref{fig:f3}, the measured correlations still show an excess over both the mapped prediction and the refitted model at small separations, most clearly at $s \leq 0.3\,h^{-1}\,\mathrm{Mpc}$ in the two lower-mass samples.

\begin{table}[tp]
\centering
\begin{tabular}{cccccc}
\hline
& $M_{h,\min}$ & $R_f$ & $a_L$ & $a_s$ & $\chi_s^2/\mathrm{dof}$ \\
\hline

& $10^{11} $ & 0.89 & 0.28 & 0.32 $\pm$ 0.03 & 2.54 \\
DMO & $10^{11.5} $ & 1.28 & 0.29 & 0.36 $\pm$ 0.03 & 1.15 \\
& $10^{12} $ & 1.84 & 0.34 & 0.34 $\pm$ 0.06 & 1.08 \\
\hline
& $10^{11} $ & 0.89 & 0.27 & 0.27 $\pm$ 0.03 & 0.94 \\
Hydro & $10^{11.5} $ & 1.27 & 0.27 & 0.32 $\pm$ 0.04 & 0.61 \\
& $10^{12} $ & 1.81 & 0.31 & 0.26 $\pm$ 0.08 & 0.95 \\
\hline

\end{tabular}
\caption{Same as table~\ref{tab:t1}, but for the redshift-space DM halo spin-axis correlations shown in figure~\ref{fig:f5}. The fitted amplitude and reduced chi-square are denoted by $a_s$ and $\chi_s^2/\mathrm{dof}$, respectively. The values of $a_L$ are taken from table~\ref{tab:t1} for comparison.} \label{tab:t2}
\end{table}

The hydrodynamic samples, shown in the bottom row of figure~\ref{fig:f5}, yield $a_s = 0.27\,\pm\,0.03$, $0.32\,\pm\,0.04$, and $0.26\,\pm\,0.08$, with corresponding reduced chi-square values of $0.94$, $0.61$, and $0.95$. These amplitudes remain comparable to their Lagrangian counterparts $a_L$, and the predictions obtained without refitting broadly follow the measured correlations. The small-scale excess seen in the DMO run is less pronounced in the hydrodynamic run. Overall, these results suggest that the separation mapping captures the general trend of the redshift-space DM spin-axis correlations.

\begin{figure}[tp]
\centering
\includegraphics[width=.98\textwidth]{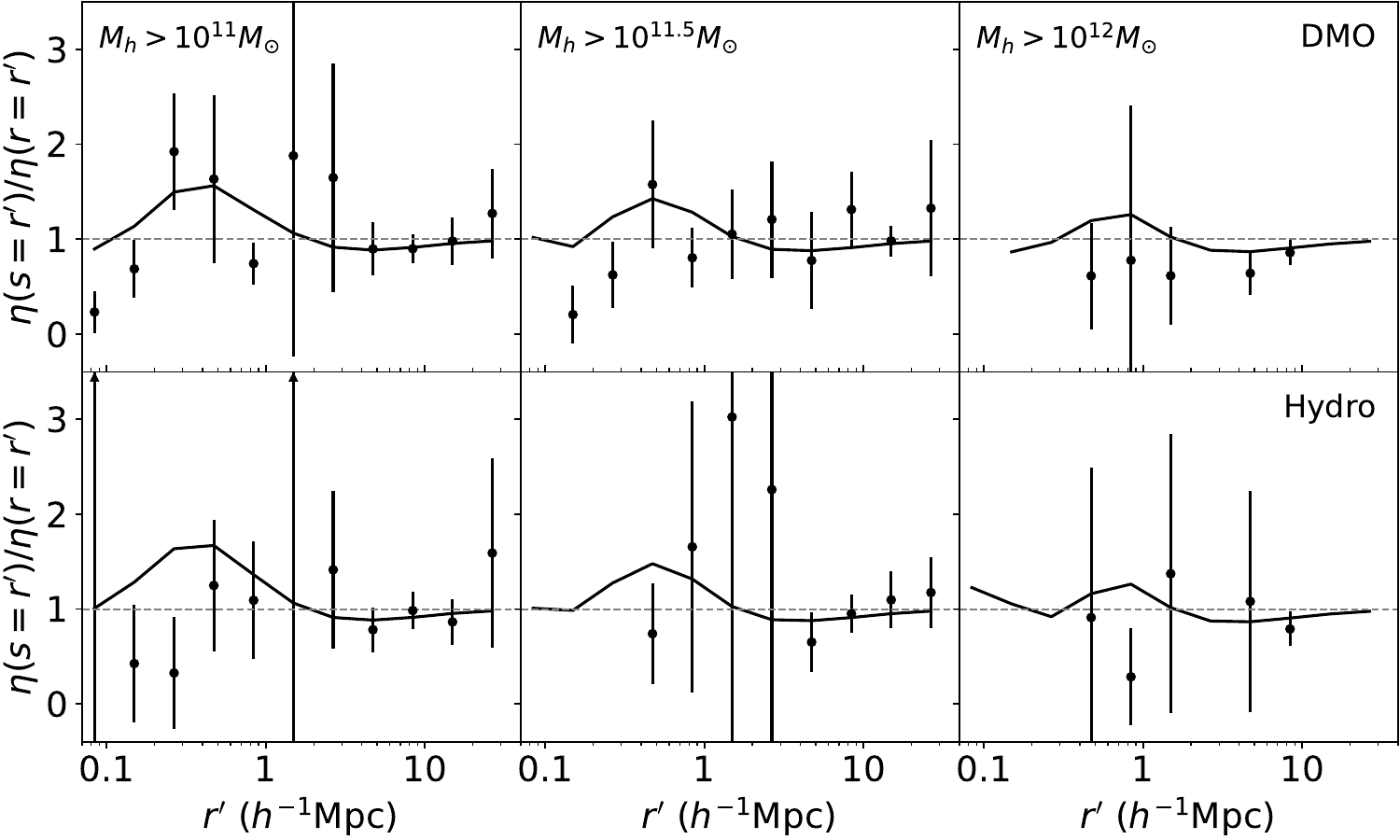}
\caption{Ratio of the redshift-space to real-space DM halo spin-axis correlation functions, $\eta(s=r')/\eta(r=r')$, evaluated at a common separation $r'$. Black points show the ratios of the measurements in figure~\ref{fig:f5} to those in figure~\ref{fig:f3}. Only bins with positive spin-axis correlations in both spaces are shown. Error bars indicate jackknife uncertainties estimated by calculating the ratio in each jackknife realization. Solid curves show the ratios of the corresponding mapped predictions using the best-fit value of $a_L$ from figure~\ref{fig:f1}. Horizontal dashed lines mark unity, and arrows indicate measurements outside the plotted range. \label{fig:f6}}
\end{figure}

Figure~\ref{fig:f6} shows the ratio of the redshift-space to real-space DM halo spin-axis correlations, $\eta(s=r')/\eta(r=r')$, evaluated at a common separation $r'$. Solid curves show the corresponding predictions obtained by mapping the same best-fit Lagrangian model to real and redshift space without refitting. The predicted ratios are similar in the DMO and hydrodynamic runs, exhibiting a modest enhancement at $r' \sim 0.3\,$--$\,1\,h^{-1}\,\mathrm{Mpc}$, followed by a mild suppression at separations of a few $h^{-1}\,\mathrm{Mpc}$ and a gradual approach to unity at larger separations. The measured ratios in the two lower-mass samples are broadly consistent with unity at $r' \geq 10\,h^{-1}\,\mathrm{Mpc}$, whereas larger variations are seen at smaller separations. In particular, the two lower-mass DMO samples show stronger suppression at the smallest separations than predicted by the mapped models, consistent with the reduced small-scale excess in redshift space seen in figure~\ref{fig:f5}. The hydrodynamic run shows substantial scatter at small separations, and the high-mass samples in both runs have large uncertainties in several bins, limiting a detailed comparison with the predicted scale dependence.

\subsection{Stellar spin-axis correlations} \label{sec:stellar}

In this section, we examine how the DM halo spin-axis correlations are reflected in the stellar component. To predict the functional form of the stellar spin-axis correlation, we consider a simple model in which the stellar spin is scattered about the corresponding halo spin with an axisymmetric distribution. We assume that this scatter is statistically independent between different subhalos and is drawn from the same distribution within each mass sample, independent of spin orientation or pair separation. Defining the misalignment angle between the DM halo and stellar spins by $\cos\theta = \hat{\boldsymbol{L}}\cdot\hat{\boldsymbol{L}}_*$, the second moment of the stellar spin component along the halo spin is $\langle\cos^2\theta\rangle$, while that along each of the two perpendicular directions is $(1-\langle\cos^2\theta\rangle)/2$. We thus obtain the following expression for the conditional second moment of the stellar spin direction:
\begin{equation}
\left\langle
\hat L_{*,i}\hat L_{*,j}
\mid\hat{\boldsymbol{L}}
\right\rangle
=
\frac{1-b}{3}\delta_{ij}
+b\hat L_i\hat L_j\,,
\end{equation}
where the stellar spin distribution about the halo spin is characterized by the parameter $b$, defined as
\begin{equation}
b\equiv\left\langle\frac{3\cos^2\theta-1}{2}\right\rangle\,.
\end{equation}

If the scatter about the halo spin directions is statistically
independent between subhalos, the conditional average over the scatter at fixed DM halo spin directions is given by
\begin{equation} \label{eqn:stellar}
\begin{aligned}
&\left\langle
\left(
\hat{\boldsymbol{L}}_*^{(1)}
\cdot\hat{\boldsymbol{L}}_*^{(2)}
\right)^2-\frac13
\,\middle|\,
\hat{\boldsymbol{L}}^{(1)},\hat{\boldsymbol{L}}^{(2)}
\right\rangle \\
&\qquad\qquad=
\left(
\frac{1-b}{3}\delta_{ij}
+b\hat L_i^{(1)}\hat L_j^{(1)}
\right)
\left(
\frac{1-b}{3}\delta_{ij}
+b\hat L_i^{(2)}\hat L_j^{(2)}
\right)
-\frac13 \\
&\qquad\qquad=
b^2\left[
\left(
\hat{\boldsymbol{L}}^{(1)}
\cdot\hat{\boldsymbol{L}}^{(2)}
\right)^2-\frac13
\right]\,.
\end{aligned}
\end{equation}
Averaging both sides of eq.~(\ref{eqn:stellar}) over pairs in each separation bin then relates the stellar and DM spin-axis correlation functions, $\eta_*(r)$ and $\eta_\mathrm{DM}(r)$:
\begin{equation}
\eta_*(r)=b^2\eta_\mathrm{DM}(r)\,.
\end{equation}
This model predicts an attenuation of the stellar spin-axis correlation while preserving the functional form of the DM spin-axis correlation, with the factor $b^2 = a_*^2/a_\mathrm{DM}^2$. 

In practice, however, predicting the attenuation factor $b^2$ is not straightforward. The model assumes that the scatter follows the same distribution within each mass sample. If the scatter depends on halo properties or environment \cite[e.g.,][]{2022ApJ...927...29L, 2023ApJ...943..128S}, a single $b^2$ may not fully describe the relative correlation amplitudes. Furthermore, the model assumes that the scatter is statistically independent between subhalos, whereas stellar spins may retain information about the initial tidal field beyond that encoded in the DM halo spins \cite{2023ApJ...952..101M}. This motivates us to provide an empirically determined effective factor, $b_{\mathrm{eff}}^2 \equiv a_{L,*}^2/a_{L,\mathrm{DM}}^2$, obtained from the best-fit amplitudes of the stellar and DM spin-axis correlations in the simulation data.

\begin{figure}[tp]
\centering
\includegraphics[width=.85\textwidth]{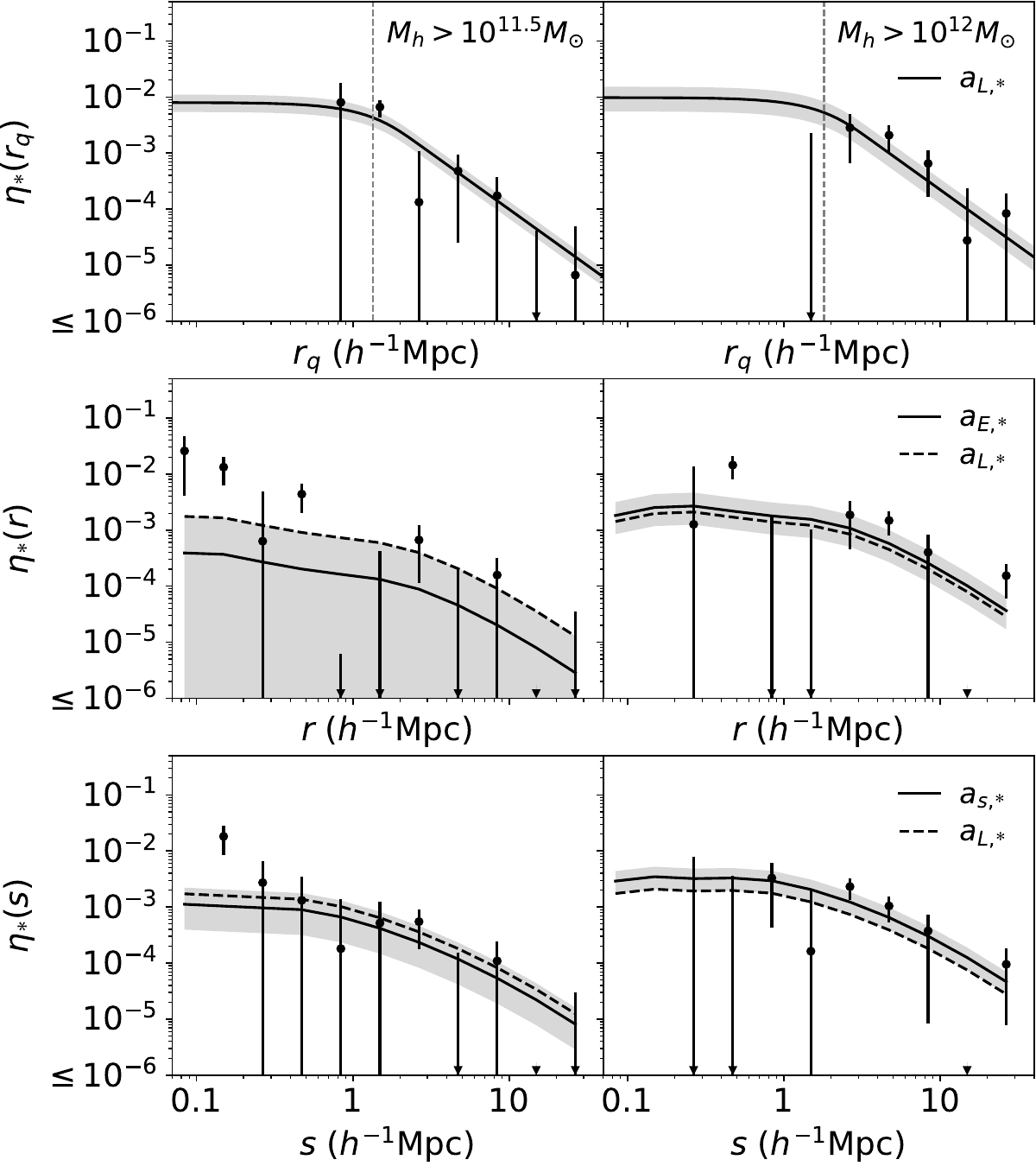}
\caption{Stellar spin-axis correlation functions, $\eta_*(r_q)$, $\eta_*(r)$, and $\eta_*(s)$, as functions of Lagrangian (top), real-space Eulerian (middle), and redshift-space Eulerian (bottom) separations. Columns show samples with lower mass limits of $10^{11.5}$ and $10^{12}\,M_\odot$, from left to right. Stellar spins are measured within twice the stellar half-mass radius, retaining only subhalos with more than 300 stellar particles within this aperture. The curves follow the conventions of figures~\ref{fig:f1}, \ref{fig:f3}, and \ref{fig:f5}, respectively, with fitted amplitudes $a_{L,*}$, $a_{E,*}$, and $a_{s,*}$. \label{fig:f7}}
\end{figure}

Figure~\ref{fig:f7} shows the stellar spin-axis correlations as functions of Lagrangian, real-space, and redshift-space separations for the samples with $M_h > 10^{11.5}\,M_\odot$ and $M_h > 10^{12}\,M_\odot$. Since most subhalos with $10^{11} < M_h/M_\odot < 10^{11.5}$ contain fewer than 300 stellar particles within $2R_{*,1/2}$, we exclude the lowest-mass sample from this analysis. The fitting results are summarized in table~\ref{tab:t3}. In Lagrangian space (top row), the LTTT-based model describes the stellar spin-axis correlations well, yielding $a_{L,*} = 0.22\,\pm\,0.04$ and $0.24\,\pm\,0.06$, with reduced chi-square values of $0.78$ and $0.89$, respectively. Compared with the DM halo spin-axis correlations in the hydrodynamic run shown in figure~\ref{fig:f1}, the stellar correlations exhibit a similar overall decline with Lagrangian separation but have somewhat lower best-fit amplitudes. The stellar correlation measurements are slightly noisier in both mass samples. In the lower-mass sample, this may be partly due to the smaller number of subhalos used for stellar spins than for DM halo spins.

\begin{table}[tp]
\centering
\setlength{\tabcolsep}{4pt}
\begin{tabular}{ccccccccc}
\hline
$M_{h,\min}$ & $R_f$ & $a_{L,*}$ & $\chi_L^2/\mathrm{dof}$ & $b_{\mathrm{eff}}^2$ & $a_{E,*}$ & $\chi_E^2/\mathrm{dof}$ & $a_{s,*}$ & $\chi_s^2/\mathrm{dof}$ \\
\hline

$10^{11.5}$ & $1.34$ & $0.22\pm0.04$ & $0.78$ & $0.60\pm0.23$ & $0.10\pm0.12$ & $1.48$ & $0.18\pm0.07$ & $0.90$ \\
$10^{12}$   & $1.81$ & $0.24\pm0.06$ & $0.89$ & $0.60\pm0.39$ & $0.27\pm0.09$ & $2.00$ & $0.31\pm0.07$ & $1.09$ \\
\hline

\end{tabular}
\caption{Same as tables~\ref{tab:t1} and \ref{tab:t2}, but for the stellar spin-axis correlations shown in figure~\ref{fig:f7}. The ratio $b_{\mathrm{eff}}^2 \equiv a_{L,*}^2/a_{L,\mathrm{DM}}^2$ compares the best-fit Lagrangian amplitudes for stellar and DM spins, with $a_{L,\mathrm{DM}}$ fitted using the same subhalo samples and smoothing scales as $a_{L,*}$. The uncertainty in $b_{\mathrm{eff}}^2$ is estimated by fitting each jackknife realization separately.} \label{tab:t3}
\end{table}

Comparing the best-fit stellar amplitudes with those of the DM spin-axis correlations for the same subhalo samples and the same smoothing scales gives $b_{\mathrm{eff}}^2 = 0.60\,\pm\,0.23$ and $0.60\,\pm\,0.39$, respectively. These values imply a moderate attenuation of the stellar spin-axis correlations relative to their DM counterparts, but are substantially larger than the attenuation factors inferred from the present-day stellar--halo spin misalignments. For our fiducial spin definitions, using all bound DM particles for the halo spins and stellar particles within $2R_{*,1/2}$ for the stellar spins, the misalignments yield $b^2 = 0.081$ and $0.062$ for the samples with $M_h > 10^{11.5}\,M_\odot$ and $M_h > 10^{12}\,M_\odot$, respectively. These values increase to $0.150$ and $0.153$ when both spins are measured using all bound particles, and to $0.356$ and $0.392$ when both are measured within $2R_{*,1/2}$. This sensitivity to the adopted spin definitions further illustrates the difficulty of inferring a single attenuation factor $b^2$ from the stellar--halo misalignment alone. The larger values of $b_{\mathrm{eff}}^2$ indicate that the measured stellar spin-axis correlations are stronger than predicted by a simple model of statistically independent stellar--halo spin misalignments, potentially reflecting additional information about the initial tidal field.

In real space (middle row), the best-fit amplitudes are $a_{E,*} = 0.10\,\pm\,0.12$ and $0.27\,\pm\,0.09$ for the lower- and higher-mass samples, respectively. The amplitude is poorly constrained for the lower-mass sample, where negative correlation measurements in several bins can pull the fit toward small amplitudes despite the clear positive correlations at small separations. For the high-mass sample, the mapped prediction and the refitted model show similar amplitudes. For both samples, the values of $a_{E,*}$ are consistent with their Lagrangian counterparts $a_{L,*}$ within the quoted $1\sigma$ uncertainties. These results suggest that the Lagrangian-to-Eulerian connection found for DM halo spins extends, at least qualitatively, to the stellar component.

We next examine the stellar spin-axis correlations in redshift space, following the procedure described in section~\ref{sec:rsd}. The results are shown in the bottom row of figure~\ref{fig:f7}. We obtain $a_{s,*} = 0.18\,\pm\,0.07$ and $0.31\,\pm\,0.07$, with reduced chi-square values of $0.90$ and $1.09$, respectively. The models reproduce the overall behavior of the stellar spin-axis correlations with separation. Relative to the predictions based on $a_{L,*}$, refitting lowers the model amplitude for the lower-mass sample and raises it for the higher-mass sample, but the quoted $1\sigma$ intervals overlap those of the Lagrangian amplitudes for both samples. Compared with the DM halo spin-axis correlations in the hydrodynamic run shown in figure~\ref{fig:f5}, the stellar correlations exhibit a similar decline with redshift-space separation, with a weaker signal in the lower-mass sample and a comparable amplitude in the higher-mass sample. Thus, although stellar spins introduce additional scatter and attenuation, their correlations can still preserve useful information about the initial tidal field at late times.

The stellar spin-axis correlations in redshift space, $\eta_*(s)$, provide a useful theoretical reference for the intrinsic spin-axis correlation accessible from observations of galaxy spins and redshifts. For comparison, ref.~\cite{2011ApJ...732...99L} obtained $a = 0.25\,\pm\,0.04$ by fitting an LTTT-based model to the observed spin-axis correlations of nearby late-type spiral galaxies in the Sloan Digital Sky Survey. This value lies between our estimates of $a_{s,*} = 0.18\,\pm\,0.07$ and $0.31\,\pm\,0.07$. A direct comparison is limited, however, by differences in the galaxy samples and by the fact that the observational analysis did not explicitly account for the separation mapping considered here. Nevertheless, the apparent similarity between our inferred amplitudes and the observational estimate is noteworthy.

\subsection{Initial spin-axis correlations} \label{sec:initial}

Our analysis has shown that the LTTT-based model, motivated by the spatial coherence of the initial tidal field, provides a useful description of the present-day spin-axis correlations when combined with the separation mapping. To complement our analysis of present-day spins and establish a more comprehensive picture, we now examine the initial DM spin-axis correlations for the same subhalo samples, using their member DM particles traced back to the initial conditions at $z = 127$. We define the initial spin vector as the angular momentum of these particles, computed analogously to eq.~(\ref{eq:spin}) using their initial positions and velocities. Comparing the initial and present-day correlations at both Lagrangian and Eulerian separations allows us to distinguish the effects of spin reorientation and pair redistribution, and to clarify how the present-day correlations relate to those established in the initial conditions.

\begin{figure}[tp]
\centering
\includegraphics[width=.98\textwidth]{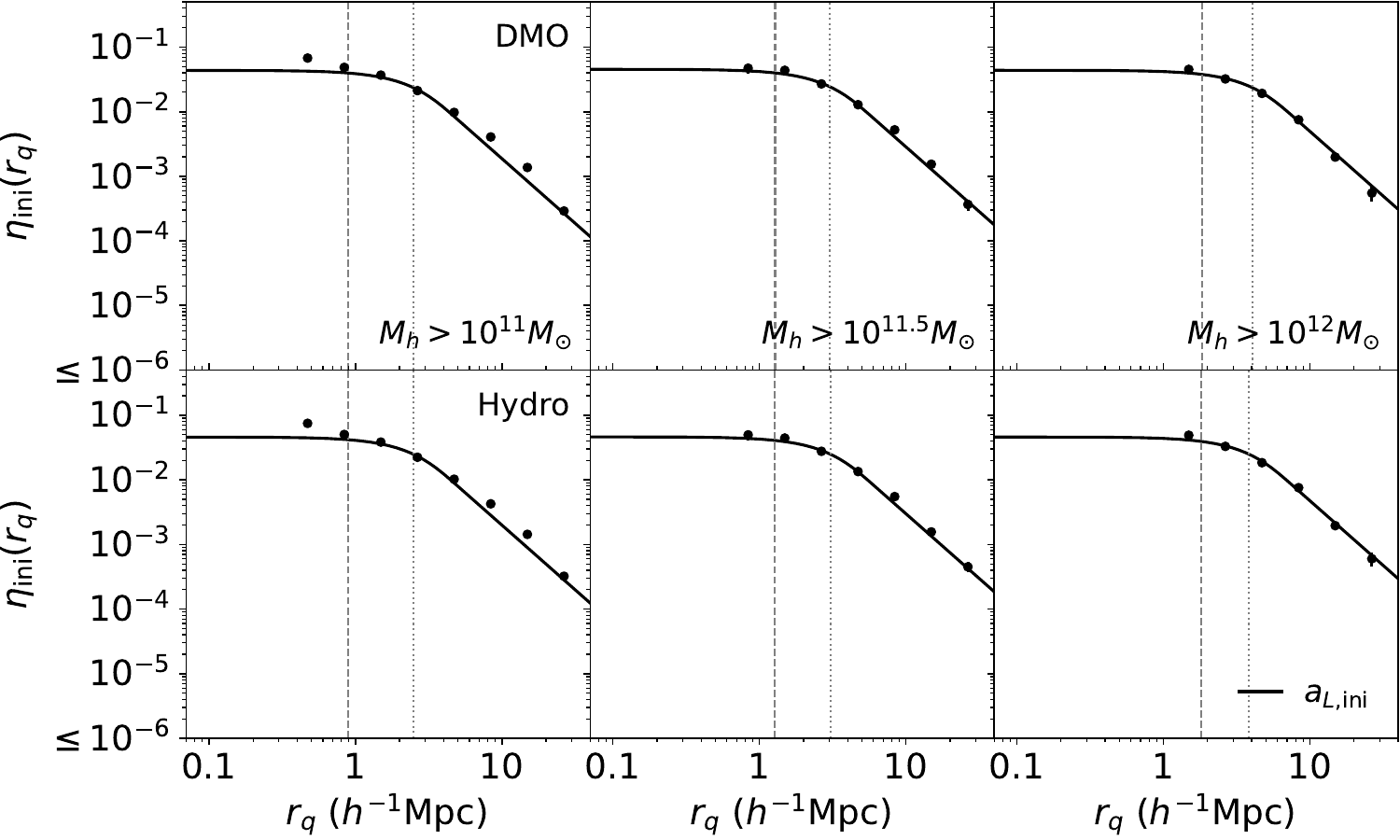}
\caption{Same as figure~\ref{fig:f1}, but for initial DM halo spin-axis correlation functions, $\eta_\mathrm{ini}(r_q)$, with both spin directions and pair separations evaluated in the initial conditions. Both $a_{L,\mathrm{ini}}$ and $R_f$ are allowed to vary in the fit. Vertical dashed and dotted lines indicate the mean Lagrangian radius $\langle R_L \rangle$ and the best-fit smoothing scale $R_f$, respectively. \label{fig:f8}}
\end{figure}

Figure~\ref{fig:f8} shows the initial DM halo spin-axis correlations, $\eta_\mathrm{ini}(r_q)$, for the six samples considered in figure~\ref{fig:f1}. All samples exhibit positive correlations that decrease with increasing Lagrangian separation, with similar behavior in the DMO and hydrodynamic runs. The initial correlations are substantially stronger than their present-day counterparts at the same $r_q$. Their relatively flat behavior extends to larger $r_q$, which the LTTT-based model fails to reproduce with $R_f$ fixed to the mean Lagrangian radius. We therefore fit these measurements using eq.~(\ref{eq:fit3d}), allowing both the amplitude $a_{L,\mathrm{ini}}$ and smoothing scale $R_f$ to vary. 

In LTTT, a protohalo acquires angular momentum through the coupling between the tidal field and the shape tensor of its extended, generally irregular Lagrangian region. A spherical smoothing window with a scale $R_f$ set solely by halo mass may therefore provide an incomplete description of the spatial structure relevant to spin-axis correlations. When measured at the same Lagrangian separations, the initial spin-axis correlations have broader profiles than those of present-day spins. This suggests that subsequent nonlinear evolution may modify the separation dependence of spin-axis correlations as well as suppress their overall amplitude, a possibility that deserves further investigation.

As summarized in table~\ref{tab:t4}, the best-fit amplitudes lie between $0.51$ and $0.53$, showing little dependence on mass threshold. The preferred smoothing scales are approximately $2.1\,$--$\,2.8$ times the mean Lagrangian radii ($R_f = 2.49\,$--$\,4.04\,h^{-1}\,\mathrm{Mpc}$), with $R_f/\langle R_L\rangle$ decreasing toward higher mass thresholds. The initial spin-axis correlations show much smaller variations among jackknife realizations than their present-day counterparts, resulting in smaller estimated uncertainties. Although the model captures the overall separation dependence, the remaining residuals are significant relative to these small uncertainties, yielding reduced chi-square values of $6.39\,$--$\,29.41$ for the two lower-mass samples in each run. The corresponding values for the highest-mass samples are $1.07$ and $1.30$ in the DMO and hydrodynamic runs, respectively.

\begin{table}[tp]
\centering
\begin{tabular}{ccccccc}
\hline
& $M_{h,\min}$ & $R_f/\langle R_L \rangle$ & $a_{L,\mathrm{ini}}$ & $\chi_L^2/\mathrm{dof}$ & $a_{E,\mathrm{ini}}$ & $\chi_E^2/\mathrm{dof}$ \\
\hline

& $10^{11} $ & 2.79 $\pm$ 0.04 & 0.51 $\pm$ 0.01 & 29.41 & 0.54 $\pm$ 0.01 & 7.75 \\
DMO & $10^{11.5} $ & 2.35 $\pm$ 0.08 & 0.52 $\pm$ 0.01 & 7.62 & 0.54 $\pm$ 0.01 & 1.77 \\
& $10^{12} $ & 2.20 $\pm$ 0.13 & 0.51 $\pm$ 0.02 & 1.07 & 0.51 $\pm$ 0.02 & 1.60 \\
\hline
& $10^{11} $ & 2.80 $\pm$ 0.05 & 0.53 $\pm$ 0.01 & 22.85 & 0.56 $\pm$ 0.01 & 9.49 \\
Hydro & $10^{11.5} $ & 2.40 $\pm$ 0.08 & 0.53 $\pm$ 0.01 & 6.39 & 0.54 $\pm$ 0.01 & 2.74 \\
& $10^{12} $ & 2.13 $\pm$ 0.15 & 0.53 $\pm$ 0.02 & 1.30 & 0.51 $\pm$ 0.02 & 1.49 \\
\hline

\end{tabular}
\caption{Same as table~\ref{tab:t1}, but for the initial DM halo spin-axis correlations shown in figures~\ref{fig:f8} and \ref{fig:f9}, with fitted amplitudes $a_{L,\mathrm{ini}}$ and $a_{E,\mathrm{ini}}$. $R_f/\langle R_L\rangle$ gives the best-fit smoothing scale normalized by the mean Lagrangian radius of each sample.} \label{tab:t4}
\end{table}

\begin{figure}[tp]
\centering
\includegraphics[width=.98\textwidth]{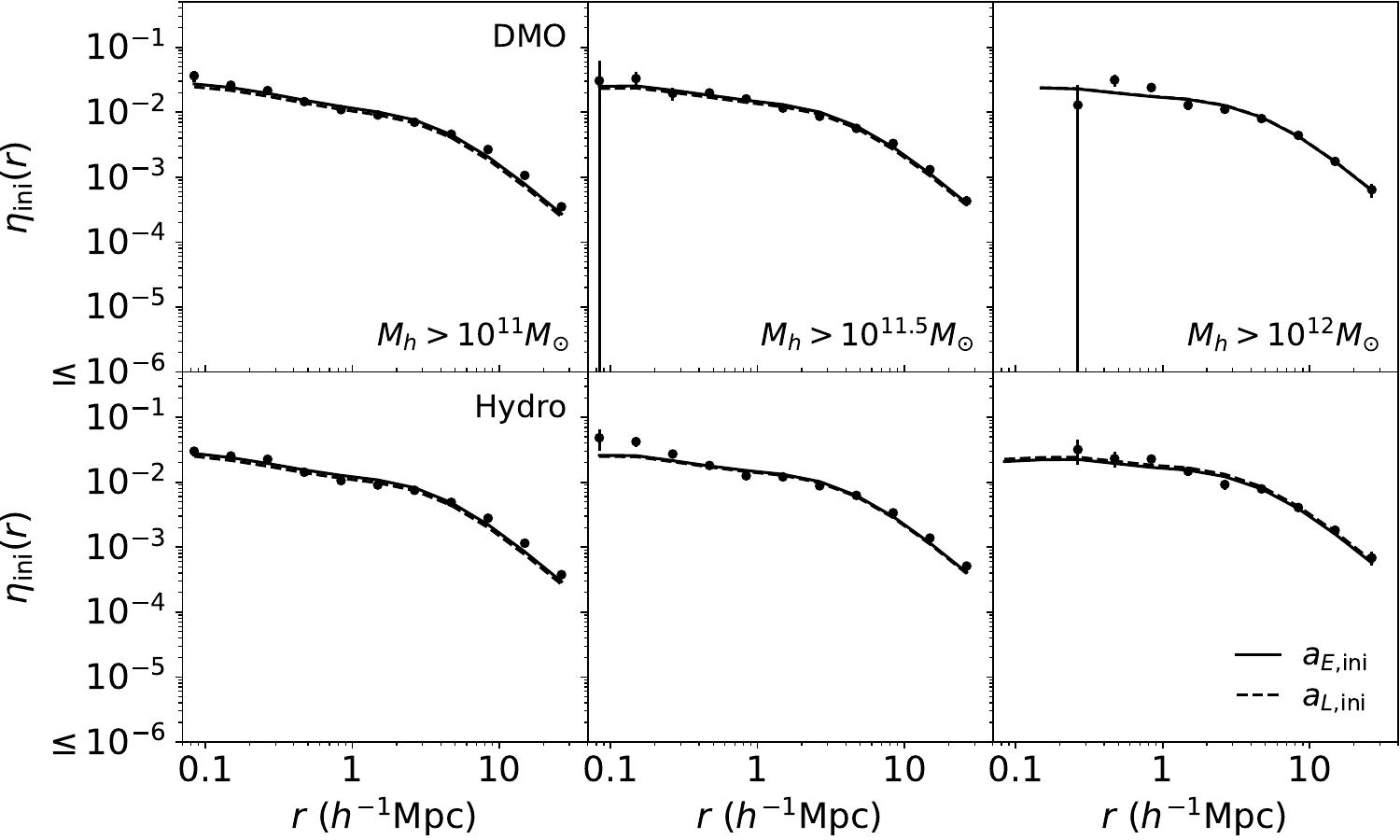}
\caption{Same as figure~\ref{fig:f3}, but for initial DM halo spin-axis correlations, $\eta_\mathrm{ini}(r)$, using the initial spin directions in figure~\ref{fig:f8} and pair separations at $z = 0$. 
The amplitude obtained by fitting directly to $\eta_\mathrm{ini}(r)$ is denoted by $a_{E,\mathrm{ini}}$, with $R_f$ fixed to the best-fit value from figure~\ref{fig:f8}. \label{fig:f9}}
\end{figure}

Figure~\ref{fig:f9} shows the correlations of the same initial spin directions as a function of present-day real-space Eulerian separation, $\eta_\mathrm{ini}(r)$. Dashed curves are obtained by applying the separation kernel $\boldsymbol{K}$ to the best-fit Lagrangian models from figure~\ref{fig:f8}, without additional fitting. Solid curves show fits to the real-space measurements with amplitude $a_{E,\mathrm{ini}}$, keeping $R_f$ fixed to its best-fit Lagrangian value. The mapped predictions capture the relatively weak decline at small separations and the steeper decrease at larger separations. Refitting changes the amplitudes only modestly, and the values of $a_{L,\mathrm{ini}}$ and $a_{E,\mathrm{ini}}$ remain close across all six samples. The reduced chi-square values nevertheless reach $7.75$ and $9.49$ for the lowest-mass samples in the DMO and hydrodynamic runs, respectively, reflecting residuals that are significant relative to the small jackknife uncertainties. These residuals cannot be attributed uniquely to the separation mapping, because the Lagrangian fits already deviate from the measured initial correlations. The overall agreement and the small changes between $a_{L,\mathrm{ini}}$ and $a_{E,\mathrm{ini}}$ support the applicability of the separation mapping to initial as well as present-day spin-axis correlations.

\begin{figure}[tp]
\centering
\includegraphics[width=.98\textwidth]{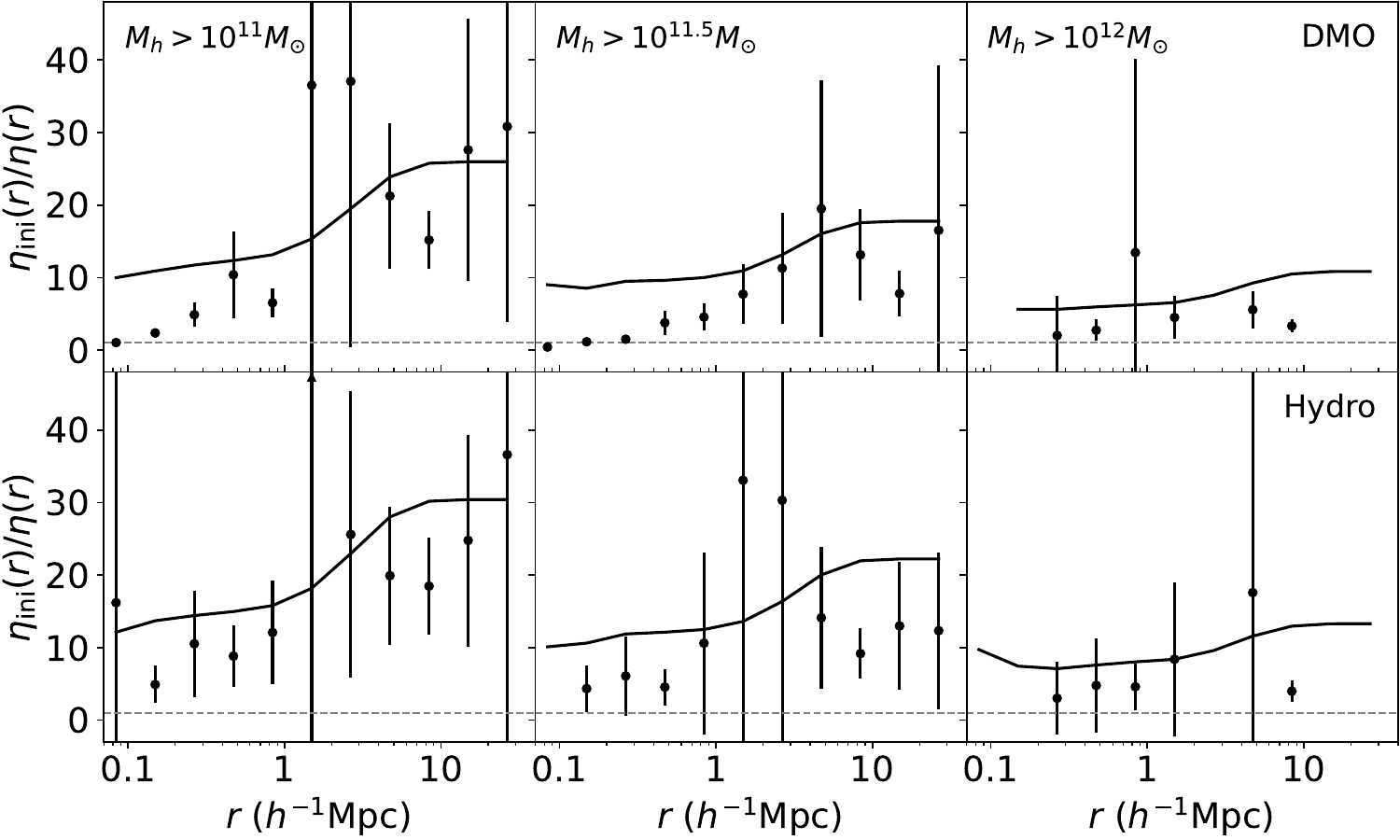}
\caption{Same as figure~\ref{fig:f6}, but for the ratio of initial to present-day DM halo spin-axis correlations, $\eta_\mathrm{ini}(r)/\eta(r)$, in real space. Black points show the ratios of the measurements in figure~\ref{fig:f9} to those in figure~\ref{fig:f3}. Solid curves show the ratios of the mapped predictions using the best-fit values of $a_L$ from figure~\ref{fig:f1} and $a_{L,\mathrm{ini}}$ from figure~\ref{fig:f8}. \label{fig:f10}}
\end{figure}

Figure~\ref{fig:f10} compares the initial and present-day DM spin-axis correlations more directly through their ratio, $\eta_\mathrm{ini}(r)/\eta(r)$, as a function of real-space Eulerian separation. The initial correlations are substantially stronger at most separations, although the ratios become increasingly uncertain at larger separations. In the DMO run, particularly for the two lower mass thresholds, the ratios are relatively small at small separations and tend to increase with separation. At the smallest separations, the present-day correlations can even be comparable to or stronger than their initial counterparts, suggesting that nonlinear spin evolution at later epochs contributes to the small-scale excess seen in figure~\ref{fig:f3}. The ratios in the hydrodynamic run are generally larger at small separations, consistent with the weaker excess in the present-day correlations. The solid curves, obtained from the mapped Lagrangian models with $a_{L,\mathrm{ini}}$ and $a_L$, broadly reproduce the relative amplitudes but overpredict the ratios at small separations. 

The overall weakening of the correlations between the initial epoch and $z = 0$ is qualitatively consistent with ref.~\cite{2002MNRAS.332..325P}, who examined the effects of nonlinear evolution on halo spin-axis correlations. Their correlations of LTTT-predicted spins remained stronger than the measured present-day correlations even after accounting for clustering evolution. Interestingly, their estimates of the amplitude parameter, $a = 0.56^{+0.03}_{-0.04}$ at $z = 50$ and $a = 0.26^{+0.05}_{-0.08}$ at $z = 0$, are comparable to our fitted values of $a_{L,\mathrm{ini}} = 0.51\,$--$\,0.53$ and $a_L = 0.27\,$--$\,0.34$, respectively. We note that the fitted values of $a_{L,\mathrm{ini}}$ are close to the idealized LTTT expectation for statistically independent inertia and tidal tensors in eqs.~(\ref{eq:spincov}) and (\ref{eq:spincov_a}).

The LTTT-based model therefore remains useful as an effective description of spin-axis correlations, despite its limitations. The statistically significant residuals in the initial correlations show that it does not fully capture their separation dependence. Moreover, different smoothing prescriptions are applied for the initial and present-day spins: $R_f$ is fixed to the mean Lagrangian radius for the present-day spins, whereas it is allowed to vary for the initial spins. Consequently, the ratio between $a_L^2$ and $a_{L,\mathrm{ini}}^2$ cannot be directly interpreted as an attenuation factor for the spin-axis correlations between the two epochs. Nevertheless, the model captures the general dependence on Lagrangian separation for both initial and present-day spins. Combined with the separation mapping obtained from the simulations, it also reproduces the broad real-space Eulerian behavior without additional fitting or an explicit model for the nonlinear evolution of spin directions.

\section{Summary and Discussion} \label{sec:summary}

\begin{figure}[tp]
\centering
\includegraphics[width=\textwidth]{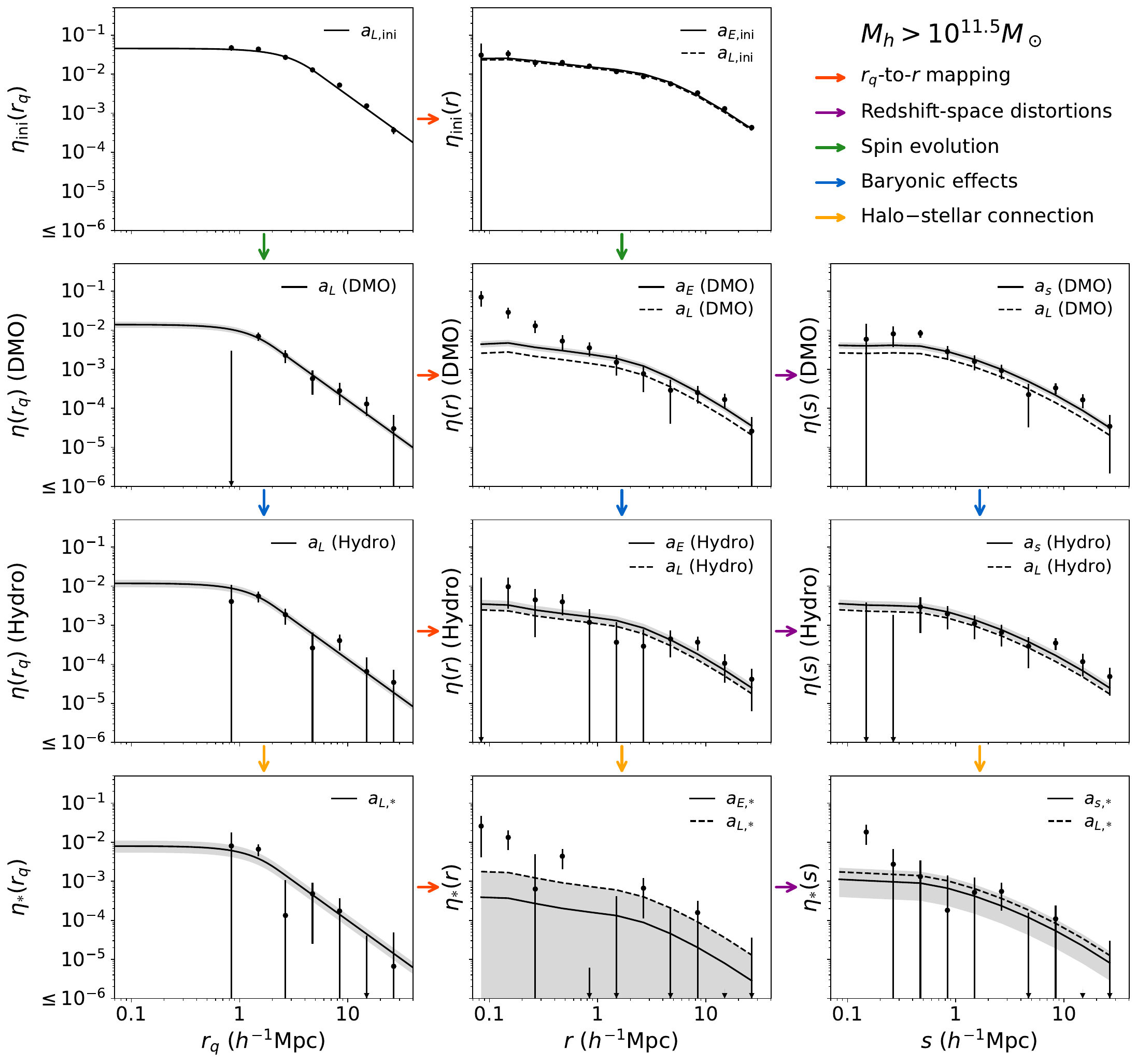}
\caption{Overview of the spin-axis correlations for the samples with $M_h > 10^{11.5}\,M_\odot$, combining selected panels from figures~\ref{fig:f1}, \ref{fig:f3}, \ref{fig:f5}, and \ref{fig:f7}--\ref{fig:f9}. Columns correspond to Lagrangian, real-space, and redshift-space separations. From top to bottom, the rows show initial DM spins in the DMO run, present-day DM spins in the DMO and hydrodynamic runs, and stellar spins in the hydrodynamic run. Symbols and curves follow the conventions of the original figures. Colored arrows indicate the connections between the measurements, as specified in the upper-right corner of the figure. \label{fig:f11}}
\end{figure}

We have used the IllustrisTNG hydrodynamic and DMO simulations to investigate whether spin-axis correlations associated with the initial tidal field persist to $z = 0$ and how they appear as a function of present-day pair separations. Figure~\ref{fig:f11} provides an overview of the results for the samples with $M_h > 10^{11.5}\,M_\odot$, highlighting the connections among the different measurements. Present-day DM spin-axis correlations are positive and decrease with Lagrangian separation, following the LTTT-based model with weak mass dependence in the fitted amplitude $a_L$ (figure~\ref{fig:f1}). Applying the conditional distribution $P(r_q \mid r)$ obtained from the simulations to the best-fit Lagrangian model largely reproduces the real-space Eulerian correlations without additional fitting, including their flatter behavior at small separations and steeper decline at larger separations (figures~\ref{fig:f2}\,--\,\ref{fig:f3}). Pair redistribution can therefore account for much of the difference between the Lagrangian and Eulerian correlation shapes. The initial DM spin-axis correlations at $z = 127$ are substantially stronger and exhibit a broader plateau, with preferred smoothing scales approximately $2.1\,$--$\,2.8$ times the mean Lagrangian radii (figure~\ref{fig:f8}). The separation mapping also captures the broad real-space Eulerian behavior of the initial spin-axis correlations (figure~\ref{fig:f9}).

The hydrodynamic and DMO runs yield similar DM spin-axis correlations at Lagrangian separations and similar separation mappings, while a clear difference appears in their small-scale real-space correlations (figures~\ref{fig:f1}\,--\,\ref{fig:f3}). The DMO run shows an excess at $r \leq 0.3\,h^{-1}\,\mathrm{Mpc}$ that is weaker in the hydrodynamic run. At the smallest separations, the present-day DMO correlations can even match or exceed their initial counterparts, suggesting that nonlinear spin evolution contributes to this excess (figure~\ref{fig:f10}). Stellar spin-axis correlations are also present at $z = 0$, with a Lagrangian separation dependence similar to that of DM spins and moderately reduced amplitudes, corresponding to $b_{\mathrm{eff}}^2 = a_{L,*}^2/a_{L,\mathrm{DM}}^2 \simeq 0.6$ for both stellar samples (figure~\ref{fig:f7}). This attenuation is substantially weaker than predicted under the assumption of statistically independent stellar--halo spin misalignments.

Using the conditional distribution $P(r_q \mid s)$, the same best-fit Lagrangian models also describe the broad redshift-space behavior of the spin-axis correlations without refitting (figures~\ref{fig:f4}\,--\,\ref{fig:f6}). The measured DM correlations remain broadly consistent between real and redshift space at separations above $10\,h^{-1}\,\mathrm{Mpc}$. The mapped models predict mild suppression at a few $h^{-1}\,\mathrm{Mpc}$ and modest enhancement around $0.3\,$--$\,1\,h^{-1}\,\mathrm{Mpc}$. At smaller separations, the measurements for the two lower-mass DMO samples show stronger suppression than predicted by the mapped models. The stellar spin-axis correlations also persist in redshift space, with refitted amplitudes consistent with their Lagrangian counterparts within the uncertainties (figure~\ref{fig:f7}). These results suggest that stellar spin-axis correlations preserve useful information about early tidal coherence and that separation mapping provides a practical connection between LTTT-based expectations and late-time measurements.

Our results share several qualitative features with previous numerical studies of halo spin correlations. Ref.~\cite{2001MNRAS.322..576H} found a weak signal for halo spin alignments at $z = 0$ at separations of $\sim\!1\,h^{-1}\,\mathrm{Mpc}$. Ref.~\cite{2002MNRAS.332..325P} also measured positive spin-axis correlations at $z = 0$, with a maximum measured amplitude of order $10^{-2}$, comparable in magnitude to the small-scale correlations in our DMO samples. Ref.~\cite{2013ApJ...762...72T} detected spin alignment at $r < 0.3\,h^{-1}\,\mathrm{Mpc}$, but no significant alignment when only the most massive subhalo in each FoF group was included. Our samples include both central and satellite subhalos and show strong correlations on these scales in the DMO run, consistent with their results, although the corresponding signal is weaker in the hydrodynamic run.

Ref.~\cite{2008ApJ...681..798L} measured positive spin-axis correlations extending to $10\,h^{-1}\,\mathrm{Mpc}$ at $z = 0$ in the Millennium simulation. Our DM spin-axis correlations likewise remain positive on comparable Eulerian scales. To describe deviations from the quadratic LTTT-based prediction, they introduced an additional term proportional to the density correlation function, motivated by the growth of non-Gaussianity in the density fluctuations. Our results show that the LTTT-based model fitted to the present-day spin-axis correlations at Lagrangian separations largely reproduces the Eulerian correlations at $z = 0$ once the mapping between the two separations is applied, without introducing an additional term. This suggests that the redistribution of halo pairs during gravitational evolution can explain much of the deviation from the quadratic prediction when the correlations are measured at Eulerian separations.

Earlier studies have already considered how changes in pair separation affect spin-axis correlations. For example, ref.~\cite{2000ApJ...543L.107P} applied a Gaussian filter to account for the effects of peculiar velocities in redshift space when comparing an LTTT-based model with observations. Ref.~\cite{2002MNRAS.332..325P} examined the effect of clustering evolution by calculating the correlations of LTTT-predicted spins as a function of present-day pair separation. Our approach extends these treatments by explicitly applying the separation mapping to the LTTT-based model using the conditional distributions $P(r_q \mid r)$ and $P(r_q \mid s)$ measured directly from the simulations. Using the best-fit Lagrangian amplitude, we can predict the correlations in both real and redshift space without additional fitting.

For stellar spins, refs.~\cite{2015MNRAS.448.3391C, 2015MNRAS.454.2736C} measured spin-axis correlations in the Horizon-AGN simulation at $z = 1.2$ and $z = 0.5$, respectively. These studies identified a dependence on galaxy color, morphology, and stellar mass, while the latter emphasized the difficulty of separating physical spin-axis correlations from grid-locking effects. Our results show that stellar spin-axis correlations are also present at $z = 0$ in IllustrisTNG, which employs a moving-mesh method rather than the adaptive mesh refinement used in Horizon-AGN. Since our samples are selected by total subhalo mass without color or morphological criteria, further work is needed to assess how these selection differences affect the stellar spin-axis correlations.

The positive stellar spin-axis correlations measured in our analysis are also qualitatively consistent with observations of nearby galaxies by refs.~\cite{2000ApJ...543L.107P, 2011ApJ...732...99L}. At higher redshifts, ref.~\cite{2025PASJ...77..389T} found no significant spin-axis correlations at angular separations corresponding to scales above $1\,h^{-1}\,\mathrm{Mpc}$ for [O\,\textsc{ii}] emitters at $z \simeq 1.2$ and $1.5$. However, a quantitative comparison with these observations requires accounting for differences in sample selection, projection effects, and the definition of the correlation estimator.

A natural next step is to apply this framework to observational galaxy samples. Integral-field spectroscopic surveys such as MaNGA \cite{2015ApJ...798....7B} and SAMI \cite{2021MNRAS.505..991C} have provided spatially resolved stellar kinematics for large galaxy samples, and ongoing surveys such as the Hector Galaxy Survey \cite{2025PASA...42..150O} are increasing the number of galaxies with such measurements. Combined with imaging information, these data allow three-dimensional galaxy spin orientations to be constrained \cite[e.g.,][]{2021MNRAS.504.4626K, 2022MNRAS.516.3569B, 2025ApJ...995L...8M, 2026MNRAS.553g2078B}. As a step toward this goal, mock observations that reproduce the sample selection and measurement effects would help connect the simulated correlations to the statistics measured from these data. These observational advances provide a timely opportunity to test the connection between early tidal coherence and observed spin-axis correlations using our framework.

Taken together, our results support the use of LTTT-based models combined with pair-separation mapping as an effective framework for interpreting late-time halo spin-axis correlations and the stellar spin-axis correlations accessible through galaxy observations. By distinguishing the evolution of spin directions from the redistribution of pairs, our analysis clarifies how early tidal coherence can be reflected in correlations measured at present-day separations. This work provides a basis for developing galaxy spin correlations as a probe of cosmological initial conditions, with their directional information complementing conventional cosmological observables.

\appendix

\section{Spin-vector correlations} \label{app:veccorr}

In this section, we examine correlations between spin vectors. Unlike the spin-axis correlation studied in the main text, the spin-vector correlation distinguishes between parallel and antiparallel spin alignments. We define the spin-vector correlation function as
\begin{equation} \label{eq:spinvector}
\eta_v \equiv
\left\langle
\hat{\boldsymbol{L}}^{(1)} \cdot \hat{\boldsymbol{L}}^{(2)}
\right\rangle\,,
\end{equation}
where the average is taken over subhalo pairs within the corresponding separation bin. We measure this statistic for both present-day and initial DM halo spins in the DMO and hydrodynamic runs, as functions of Lagrangian and real-space Eulerian separations.

\begin{figure}[tp]
\centering
\includegraphics[width=.98\textwidth]{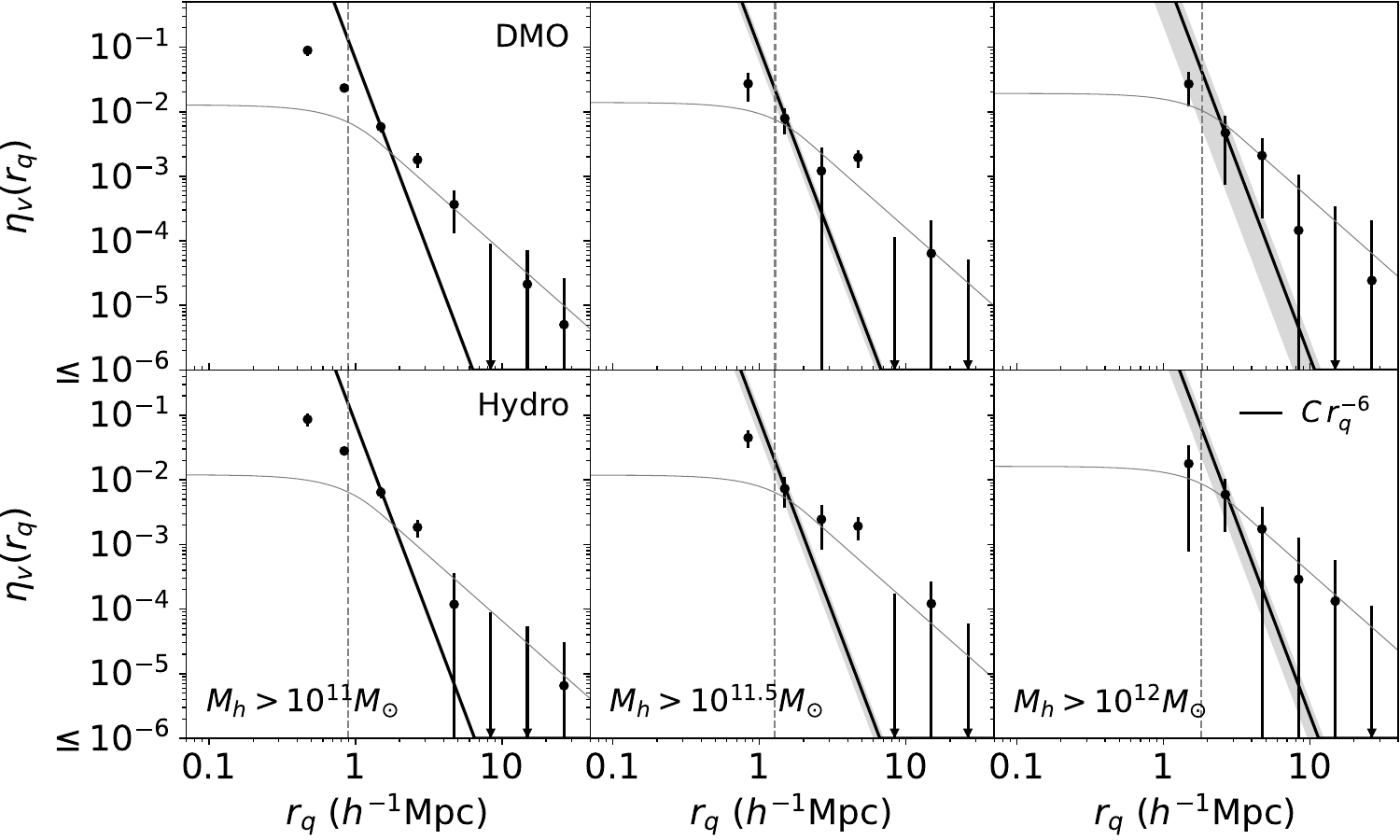}
\caption{Same as figure~\ref{fig:f1}, but for DM halo spin-vector correlation functions, $\eta_v(r_q)$, as a function of Lagrangian separation $r_q$. Black solid curves show the best-fit power-law model, $Cr_q^{-6}$, fitted over $r_q > \langle R_L\rangle$ (shown as vertical dashed lines). Shaded regions indicate the $1\sigma$ uncertainties of the fitted models. In this and subsequent figures, thin gray curves show the best-fit spin-axis correlation models for comparison. \label{fig:f12}}
\end{figure}

\begin{figure}[tp]
\centering
\includegraphics[width=.98\textwidth]{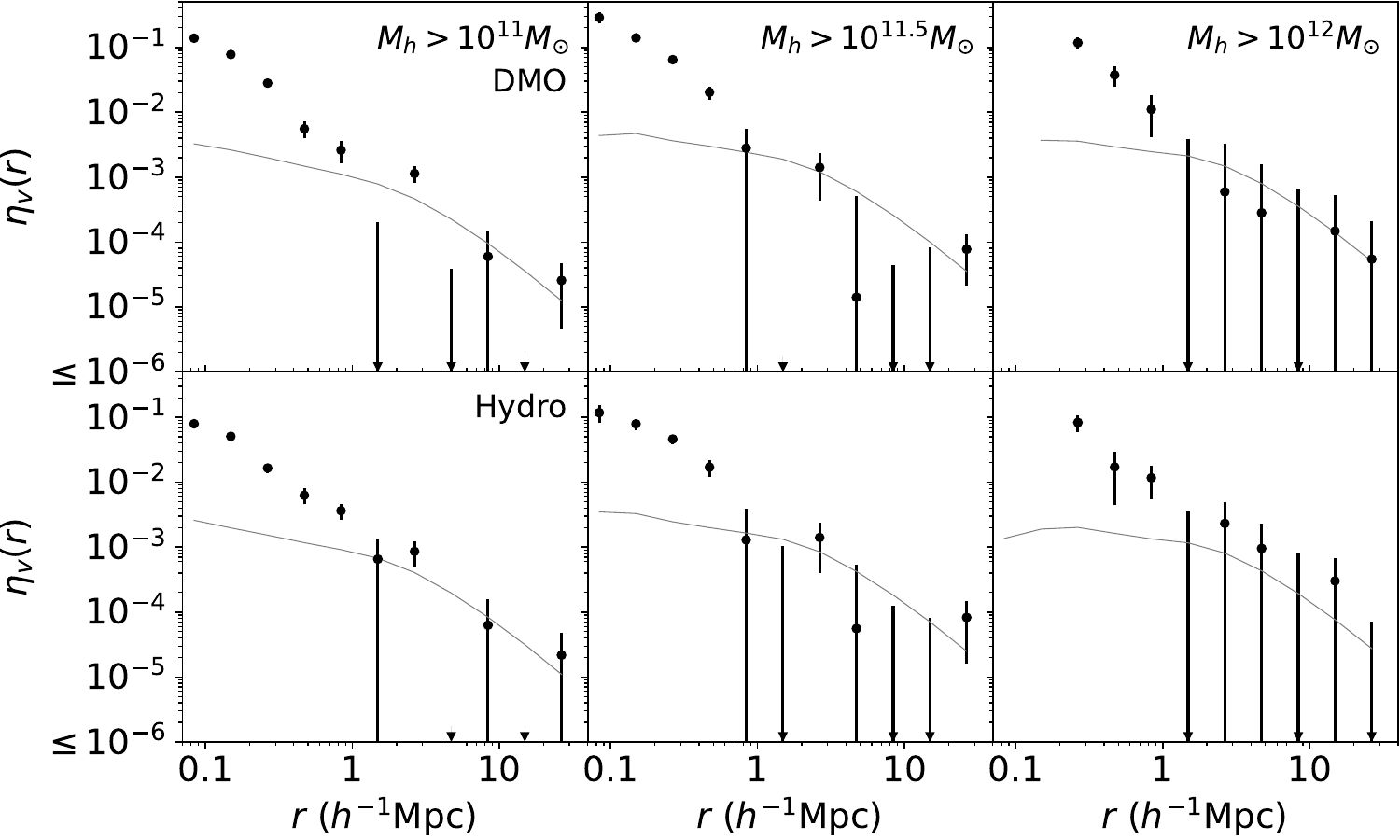}
\caption{Same as figure~\ref{fig:f3}, but for DM halo spin-vector correlation functions, $\eta_v(r)$, as a function of real-space Eulerian separation $r$. \label{fig:f13}}
\end{figure}

Figures~\ref{fig:f12} and \ref{fig:f13} show the present-day DM halo spin-vector correlations as functions of Lagrangian and real-space Eulerian separations, respectively. Thin gray curves show the best-fit models for the corresponding spin-axis correlations. In both runs, the correlations are positive at small separations and decrease toward zero with increasing separation. As in the spin-axis correlations, the DMO and hydrodynamic runs show similar trends in Lagrangian space, while the hydrodynamic run exhibits slightly lower amplitudes at the smallest real-space separations. The spin-vector correlations, however, increase much more sharply toward small separations, whereas the spin-axis models become relatively flat. This difference is particularly evident in the lowest-mass sample in figure~\ref{fig:f12} and across all three samples in figure~\ref{fig:f13}. At larger separations, the spin-vector correlations exhibit larger jackknife uncertainties than the corresponding spin-axis correlations.

\begin{figure}[tp]
\centering
\includegraphics[width=.98\textwidth]{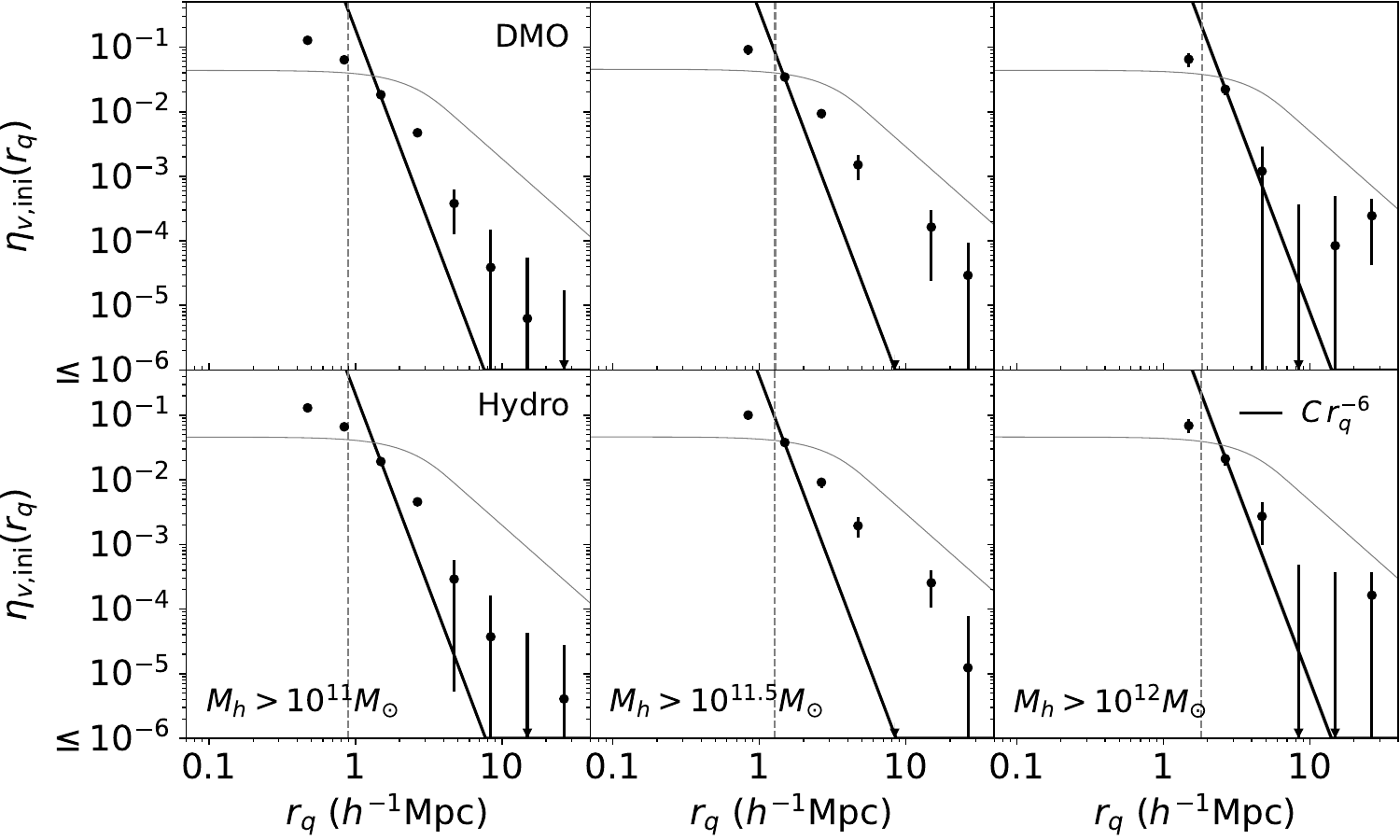}
\caption{Same as figure~\ref{fig:f8}, but for initial DM halo spin-vector correlation functions, $\eta_{v,\mathrm{ini}}(r_q)$, as a function of Lagrangian separation $r_q$. Black solid curves show the best-fit power-law model, $Cr_q^{-6}$, fitted over $r_q > \langle R_L\rangle$ (shown as vertical dashed lines). Shaded regions indicate the $1\sigma$ uncertainties of the fitted models. \label{fig:f14}}
\end{figure}

\begin{figure}[tp]
\centering
\includegraphics[width=.98\textwidth]{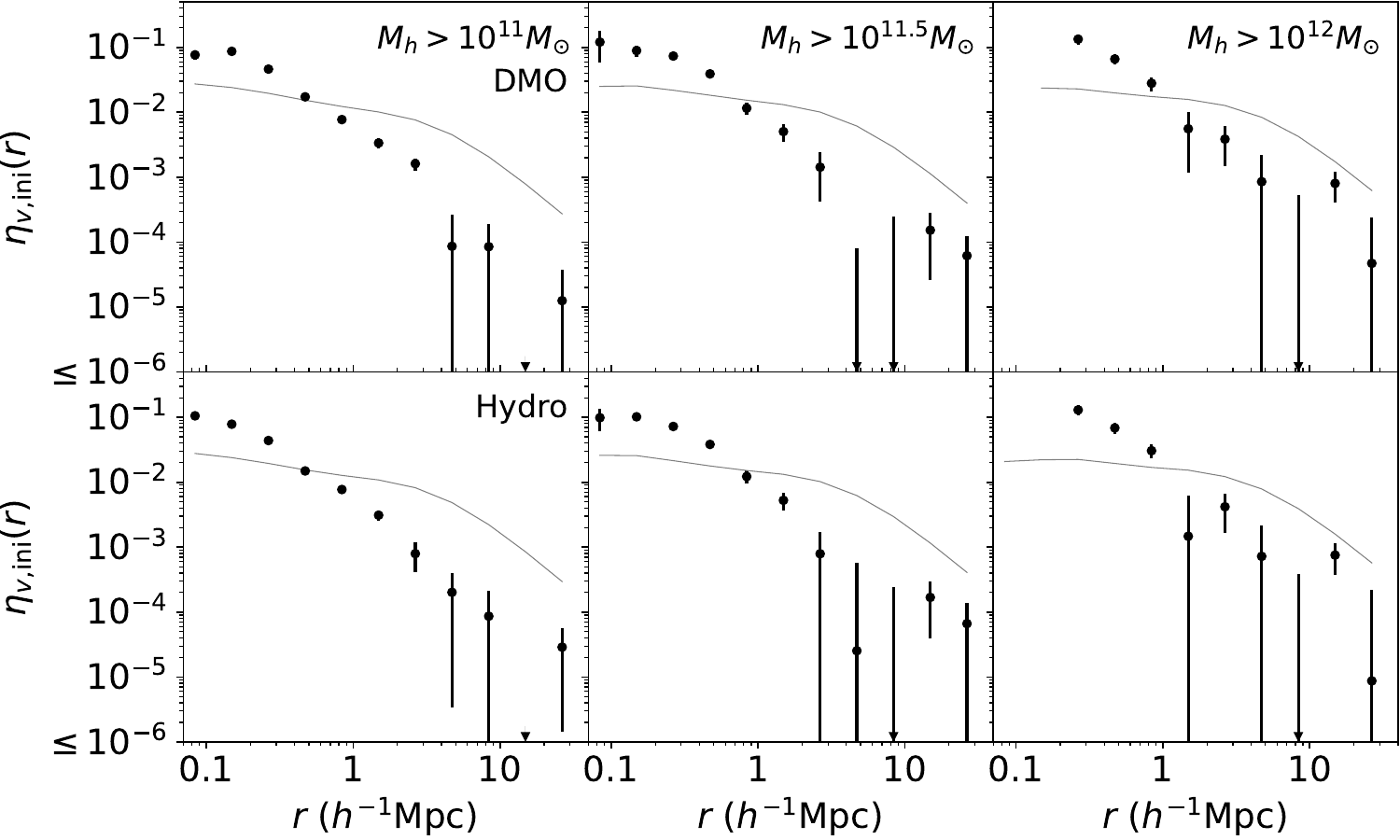}
\caption{Same as figure~\ref{fig:f9}, but for initial DM halo spin-vector correlation functions, $\eta_{v,\mathrm{ini}}(r)$, as a function of present-day real-space Eulerian separation $r$. \label{fig:f15}}
\end{figure}

Figures~\ref{fig:f14} and \ref{fig:f15} show the corresponding results for the initial DM halo spins, using Lagrangian and present-day real-space Eulerian separations, respectively. The DMO and hydrodynamic runs again exhibit very similar behavior. The correlations are positive at small separations and decline more steeply with increasing separation than the initial spin-axis models shown for reference. They generally fall below these models at separations larger than $1\,$--$\,2\,h^{-1}\,\mathrm{Mpc}$. The presence of spin-vector correlations in the initial spins indicates that they cannot be attributed solely to subsequent nonlinear spin evolution.

In the LTTT-based model adopted in section~\ref{sec:model}, opposite spin directions are equally probable at fixed tidal tensors. If the inertia tensors of different subhalos are statistically independent given the tidal tensors, the spin-vector correlation vanishes even when the spin-axis correlation is nonzero \cite{2009MNRAS.392.1225S}. Explaining the nonzero spin-vector correlations within LTTT therefore requires accounting for spatial correlations involving both the inertia and tidal tensors. To estimate the resulting separation dependence, we adopt an approximation in which the Hessian of the smoothed initial density field is used as a proxy for the protohalo inertia tensor \cite{2015MNRAS.452.3369C, 2021NatAs...5..283M}. Under this approximation, the spin vector can be expressed as $J_i \propto \epsilon_{ijk}H_{j\ell}T_{\ell k}$, where $H_{ij} \equiv \partial_i\partial_j\delta_R$ is the Hessian of the smoothed density field and $T_{ij}$ is the tidal tensor. 

For a Gaussian initial density field, applying Wick's theorem to the unnormalized spin-vector correlation, written schematically as $\langle\boldsymbol{J}\cdot\boldsymbol{J}'\rangle\propto\langle HTH'T'\rangle$, gives three pairings: $\langle HT\rangle\langle H'T'\rangle$, $\langle HH'\rangle\langle TT'\rangle$, and $\langle HT'\rangle\langle TH'\rangle$. The first term is proportional to $\langle\boldsymbol{J}\rangle\cdot\langle\boldsymbol{J}'\rangle$ and therefore vanishes in a statistically isotropic density field. For a power-law matter power spectrum, $P_\delta(k)\propto k^{-2}$, the tensor correlations at separations much larger than the smoothing scale behave as $\langle TT'\rangle\propto r_q^{-1}$, $\langle HT'\rangle\propto r_q^{-3}$, and $\langle HH'\rangle\propto r_q^{-5}$, so the latter two terms both scale as $r_q^{-6}$. The analytic models for unit spin vectors, however, are more difficult to evaluate because normalization introduces the denominator $|\boldsymbol{J}||\boldsymbol{J}'|$, which depends on the same tensors as the numerator. At large separations, where the spatial correlations between the underlying tensors are weak, we may assume that normalization preserves the leading separation dependence, giving $\eta_v(r_q)\propto r_q^{-6}$, which is much steeper than the $r_q^{-2}$ scaling predicted for the spin-axis correlation under the same power-law assumption.

To compare this prediction with the measured spin-vector correlations, we fit a power-law model, $Cr_q^{-6}$, over the range $r_q > \langle R_L\rangle$, allowing only the amplitude $C$ to vary. As described in section~\ref{sec:methods}, the fits use $\chi^2$ minimization with the corresponding full jackknife covariance matrix. The best-fit models are shown as black solid curves in figures~\ref{fig:f12} and \ref{fig:f14}. The steeper decline of the spin-vector correlations relative to the spin-axis correlations is qualitatively consistent with the above prediction. However, for both present-day and initial spins, the measured correlations decline less steeply than $r_q^{-6}$ over the fitting range. This discrepancy may reflect limitations of the assumption of Gaussianity in the Wick expansion, the local density Hessian approximation, and our simplifying assumptions about the power spectrum and spin normalization. We leave more detailed modeling of spin-vector correlations for future work.

The relatively strong signal at small separations suggests that spin-vector correlations may be detectable observationally. Interestingly, they remain strong compared with spin-axis correlations even for present-day spins, as shown in figure~\ref{fig:f13}. However, their steeper decline with separation and larger statistical uncertainties compared with spin-axis correlations may make their detection particularly challenging at separations beyond $1\,$--$\,2\,h^{-1}\,\mathrm{Mpc}$.

\acknowledgments
J.-S.M. acknowledges support from the Academia Sinica Postdoctoral Scholar Program under grant No. AS-PD-1152-M01-1. T.O. acknowledges support from the Taiwan National Science and Technology Council under grant No. NSTC 114-2112-M-001-004- and No. NSTC 115-2112-M-001-059-MY3, and the Academia Sinica Investigator Project grant No. AS-IV-114-M03 for the period of 2025--2029. U.-L.P. is supported by the Natural Sciences and Engineering Research Council of Canada (NSERC) (funding reference number RGPIN-2019-06770, ALLRP 586559-23, RGPIN-2025-06396), the Ontario Research Fund -- Research Excellence (Project No. RE012-045, 72074697), AMD AI Quantum Astro (Fund No. 473735), the Academia Sinica (Grand Challenge Seed Program AS-GCS-114-M02 and Investigator Award AS-IV-115-M04), and the National Science and Technology Council (NSTC) of Taiwan (Grant No. 115-2123-M-001-009).

The IllustrisTNG simulations were undertaken with compute time awarded by the Gauss Centre for Supercomputing (GCS) under GCS Large-Scale Projects GCS-ILLU and GCS-DWAR on the GCS share of the supercomputer Hazel Hen at the High Performance Computing Center Stuttgart (HLRS), as well as on the machines of the Max Planck Computing and Data Facility (MPCDF) in Garching, Germany. The authors used ChatGPT (GPT-6) and GitHub Copilot to assist with manuscript preparation, language editing, mathematical derivations, and code development. All AI-assisted content was carefully reviewed, verified, and edited by the authors, who take full responsibility for the final content of the manuscript.


\bibliographystyle{JHEP}
\bibliography{biblio.bib}

\end{document}